\documentclass[11pt]{article}
\usepackage{a4}
\usepackage{amsmath}
\usepackage{amsfonts}
\newcommand{\vc}[1]{{\mathbf{#1}}}
\newcommand{\mx}[1]{\vc{#1}}
\newcommand{\be}{\begin{equation}}
\newcommand{\ee}{\end{equation}}
\newcommand{\bea}{\begin{eqnarray}}
\newcommand{\eea}{\end{eqnarray}}

\renewcommand{\d}{\mathrm{d}}

\newcommand{\dd}[2][]{\ensuremath{\frac{\d #1}{\d #2}}}

\newcommand{\Ra}{\Rightarrow}

\DeclareMathSymbol{\mg}{\mathrel}{symbols}{"1D}
\newcommand{\ml}{\ll}

\newcommand{\Bnabla}{\ensuremath{\boldsymbol{\nabla}}}
\newcommand{\Bder}{\ensuremath{\boldsymbol{\partial}}}

\newcommand{\ga}{\alpha}
\newcommand{\gb}{\beta}
\renewcommand{\gg}{\gamma}
\newcommand{\gd}{\delta}
\renewcommand{\ge}{\epsilon}

\newcommand{\gf}{\phi}

\newcommand{\gm}{\mu}

\newcommand{\gk}{\kappa}
\newcommand{\gl}{\lambda}
\newcommand{\gr}{\rho}
\newcommand{\gth}{\theta}
\newcommand{\gs}{\sigma}
\newcommand{\gt}{\tau}
\newcommand{\go}{\omega}

\newcommand{\gp}{\pi}
\newcommand{\gps}{\psi}
\newcommand{\get}{\eta}
\newcommand{\gch}{\chi}
\newcommand{\gG}{\Gamma}
\newcommand{\gD}{\Delta}
\newcommand{\gF}{\Phi}

\newcommand{\gL}{\Lambda}

\newcommand{\gTh}{\Theta}

\newcommand{\gPs}{\Psi}

\newcommand{\cD}{{\cal D}}

\newcommand{\cH}{{\cal H}}

\newcommand{\cL}{{\cal L}}
\newcommand{\cM}{{\cal M}}

\newcommand{\cO}{{\cal O}}

\newcommand{\tA}{{\tilde A}}

\newcommand{\tC}{{\tilde C}}

\newcommand{\tF}{{\tilde F}}

\newcommand{\tr}{\text{tr}}

\newcommand{\Id}{\text{\small 1}\hspace{-3.5pt}\text{1}}

\newcommand{\lra}{\longrightarrow}

\newcommand{\ra}{\rightarrow}

\newcommand{\der}{\partial}

\newcommand{\inv}{^{-1}}
\newcommand{\nit}{\noindent}

\newcommand{\np}{\newpage}
\newcommand{\dsp}{\displaystyle}

\newcommand{\lh}{\left(}
\newcommand{\rh}{\right)}

\newcommand{\labl}[1]{\label{#1}}
\newcommand{\half}{\frac 12 }

\newcommand{\beq}{\begin{equation}}
\newcommand{\eeq}{\end{equation}}
\newcommand{\barr}{\begin{array}}
\newcommand{\earr}{\end{array}}
\newcommand{\equ}[1]{\begin{gather} #1 \end{gather}}
\newcommand{\equa}[1]{\begin{align} #1 \end{align}}

\newcommand{\mtrx}[1]{\begin{matrix} #1 \end{matrix}}
\newcommand{\pmtrx}[1]{\begin{pmatrix} #1 \end{pmatrix}}

\newcommand{\non}{\nonumber}
\newcounter{oldcounter}

\newcommand{\tge}{{\tilde\epsilon}}

\newcommand{\tgf}{{\tilde\phi}}

\newcommand{\tgx}{{\tilde\xi}}

\newcommand{\tget}{{\tilde\eta}}

\newcommand{\Bgd}{{\boldsymbol \delta}}

\newcommand{\Bgf}{{\boldsymbol \phi}}

\newcommand{\Bgx}{{\boldsymbol \xi}}

\newcommand{\Bget}{{\boldsymbol \eta}}

\newcommand{\Real}{\mathbb{R}}

\newcommand{\Bsfm}{\mbox{{\sffamily\bfseries m}}}   
\newcommand{\qand}{\quad \text{and} \quad}

\begin{document}

\pagestyle{empty}

\begin{flushright}
SPIN-2000/31\\
ITP-UU-00/35\\
NIKHEF 00-038\\
hep-ph/0011325

\end{flushright}

\begin{center}
{\Large {\bf 
Density perturbations arising from multiple field slow-roll inflation}} \\
\vspace{5ex}

{\large S.\ Groot Nibbelink$^1$ and B.J.W.\ van Tent$^2$}\\
\vspace{3ex}
{\em $^1$Physikalisches Institut, Universit\"at Bonn} \\
{\em Nu\ss allee 12, D-53115 Bonn, Germany} \\
{Email: nibblink@th.physik.uni-bonn.de}\\
\vspace{3ex}
{\em $^2$Spinoza Institute/Institute for Theoretical Physics, Utrecht University}\\
{\em P.O.\ Box 80195, 3508 TD Utrecht, The Netherlands}\\
{Email: B.J.W.vanTent@phys.uu.nl}\\

\vspace{5ex}

December 13, 2000

\vspace{15ex}

\end{center}
{\small 
{\bf Abstract}

In this paper we analyze scalar gravitational perturbations on a
Robertson-Walker background in the presence of multiple scalar fields that take
values on a (geometrically non-trivial) field manifold during slow-roll
inflation. For this purpose modified and generalized slow-roll functions are
introduced and their properties examined. These functions make it possible to
estimate to what extent the gravitational potential decouples from the scalar
field perturbations. The correlation function of the gravitational potential is
calculated in an arbitrary state. We argue that using the vacuum state seems a
reasonable assumption for those perturbations that can be observed in the CMBR.
Various aspects are illustrated by examples with multiple scalar fields that
take values on flat and curved manifolds. 

}

\vspace{1cm}

\nit {\em PACS number:} 98.80.Cq

\nit {\em Keywords:} slow-roll inflation; density perturbations; non-trivial
scalar  field geometry.

\np

\pagestyle{plain}
\pagenumbering{arabic}

\section{Introduction}

As has been known for a long time, inflation  \cite{Guth,boekLinde} offers a
mechanism for the production of density perturbations,  which are supposed to
be the seeds for the formation of large scale structures in the universe.  This
mechanism is the magnification of microscopic quantum fluctuations in the
scalar fields present during the inflationary epoch into macroscopic matter and
metric perturbations. Also, since a part of the primordial spectrum of density
perturbations is observed in the cosmic microwave background radiation (CMBR),
this mechanism offers one of the most important ways of checking and
constraining possible models of inflation, see e.g.\ \cite{Kinneyetal}.

The theory of the production of density perturbations in the case of a single
real scalar field has been studied for a long time
\cite{Bardeenetal,Mukhanovetal,LiddleLyth}. However, to realize inflation that
leads to the observed density perturbations in a model without very unnatural
values of the parameters, it is now thought that one needs more than one field.
This is a strong motivation for hybrid inflation models \cite{Linde} (more
models can be found in \cite{LythRiotto}). Also, many high-energy theories
contain a lot of scalar fields.  The Higgs sector of the standard model
consists of  one physical particle, but in grand unification or supersymmetric
models one  expects many more scalars. Ultimately one would hope to be able to 
identify those fields that could act as inflatons. For all these reasons it is
important to develop a theory for perturbations from multiple field inflation
as well. Work in this direction has been done by several people.  Using gauge
invariant variables the authors of \cite{NambuTaruya,Gordonetal} treated two
field inflation. The fluid flow approach was extended to multiple fields in 
\cite{LythRiotto}, while a more geometrical approach was used in
\cite{SasakiStewart,NakamuraStewart}; both methods assumed several
slow-roll-like conditions on the potential. Using slow-roll approximations
for both the background and the perturbation equations the authors of
\cite{MukhanovSteinhardt,PolarskiStar} were able to find expressions for the 
metric perturbations in multiple field inflation.

%The main ingredient of inflation is the presence of an effective cosmological 
%constant. This could be due to the potential of a scalar field theory, if its
%slope is small enough so that the scalar fields move downwards only very slowly
%(slow-roll inflation). 
In this paper we generalize the slow-roll parameters for a single background
field to the multiple field case in a systematic way. We can then give a clear 
quantification of the relative importance of terms in the equations obtained by
extending the single field density perturbation calculations by Mukhanov, 
Feldman and Brandenberger \cite{Mukhanovetal} to multiple fields. In our
definition of the slow-roll functions we do not implicitly assume slow roll to
be valid. As a consequence, these slow-roll functions are expressed in terms of
derivatives of the field velocity and the Hubble parameter, but not the
potential as the conventional slow-roll parameters are. A big advantage of this
is that these slow-roll functions can be identified in all kinds of equations
that are still exact, which is the reason why we can estimate  the relative
importance of various terms and make a well-motivated decision about neglecting
some of them.  In the case of multiple scalar field inflation it is very
convenient to think in  terms of vectors: if the fields are local coordinates
on a curved manifold, their derivatives and their fluctuations can be
interpreted as covariant vectors  in the tangent bundle of the manifold. Since
the fluctuations are assumed to be  small, a linearization procedure can be
used to obtain equations for the  perturbations. In these equations  a
prominent role is played by the scalar field  velocity. Since in single scalar
field inflation the only direction is parallel to the field velocity, it is to
be expected that the parallel field perturbations can be absorbed in the
gravitational perturbations, as happens in the single field theory. Using the
modified definitions  of the slow-roll functions we can investigate to what
extent this is correct. 

In the CMBR spectrum we can observe correlations in the temperature 
distribution. They are assumed to be due to gravitational perturbations that 
are of a quantum origin at the beginning of inflation. To calculate these 
correlations we therefore need to address the question of what is the relevant
state at the initial stages of inflation. We argue that
the  conventionally used vacuum state seems a good assumption. We investigate
the effect on the correlator in the (probably unrealistic) case that the true 
state at the beginning of inflation is thermal with a temperature of the order 
of the Planck scale. For most of this work we restrict ourselves to the
calculation of the correlation function near the end of inflation. However, to 
show that our results are consistent with other  results  in the literature,
the correlation function is also evaluated at the time of recombination, 
ignoring possible problems during (p)reheating. 

During inflation there is a relatively sharp transition in the behaviour of a
fluctuation when the corresponding wavelength `passes through the horizon'
(this is discussed in detail in this paper). This moment of horizon crossing
can be used to identify a certain  scale $k$. The smallest scale (or largest
wavelength) that can be observed in the CMBR is the one that is reentering the
horizon at this very moment,  indicated by $k_0$. Assuming that the
fluctuations do not change once they have passed outside the
horizon,\footnote{In \cite{Wandsetal} it was proved that this is always true
for the so-called adiabatic perturbations, but it need not be true for the
so-called isocurvature perturbations, see \cite{FinelliBrand}. A full
treatment of this complication is beyond the scope of  this paper.} we can
observe the undisturbed inflationary perturbation spectrum up to the scale that
reentered the horizon at the time of recombination when the CMBR was formed,
which has $\ln(k_{\mathrm{rec}}/k_0) \approx 3.5$. On the other hand, larger
scales already reentered the horizon before this time, and so on these scales
the spectrum has been influenced by physical processes taking place long after
inflation. The largest scale expected to be measured with the Planck satellite
has  $\ln(k/k_0) \approx 7$ (corresponding with multipole $l=2000$).  An
important quantity in our discussions is the number of e-folds $N_k$ that occur
after a certain  scale $k$ crosses the horizon during inflation, until the end
of inflation. As is derived in e.g.\ equation (5.16) of \cite{LiddleLyth},
$N_k$ depends logarithmically on $k$, taking a value of about 60 for $k_0$.
This number has only a logarithmic dependence on model-dependent quantities
like the reheating temperature. Hence the observationally important scales
cross the horizon about 50 to 60 e-folds before the end of inflation.

Apart from this introduction and the conclusions at the end,  
the paper is structured as follows. 
In section \ref{scalpert} the theory of scalar perturbations with a
gravitational potential coupled to a single scalar field is reviewed, 
following the methods discussed in \cite{Mukhanovetal}. 
This allows us to introduce various relevant concepts. 
In addition, this section also makes it possible to compare 
the single and multiple scalar field situations 
in the following sections.

The generalization to multiple scalar field inflation is developed in  section
\ref{multi}. As these scalar fields parameterize a possibly  curved manifold,
it is necessary and convenient to introduce some  geometrical tools in
subsection \ref{intermezzo}. Two examples   of these are the inner product and
the covariant derivatives  associated with the metric of the manifold.  Another
concept introduced here is the projection on directions parallel and 
perpendicular to the background field velocity, which  plays an important role
in the decoupling of the gravitational perturbations from the independent
scalar field fluctuations. In the next subsection we explain why the single
field method does not lead to an immediate decoupling of the gravitational
perturbations in the multiple field case.
To be able to compare the relative size of terms in the equations
we define  modified versions of the slow-roll functions that can be used in
multiple field inflation in subsection \ref{slowrollsec} and derive various 
useful properties for them. 
In subsection \ref{Pertsolv} we derive the equation of motion for the
gravitational potential in terms of these slow-roll functions and show that the
term coupling this equation to the perpendicular field perturbations is small,
so that also in the multiple field case there is an effective decoupling.
The equation of motion for the perpendicular scalar field perturbations is 
derived in the appendix using similar methods.
A discussion of the solution for the decoupled gravitational potential 
concludes this subsection. 

Section \ref{spectrum} is devoted to the computation of the quantum 
correlation function of the gravitational potential. We argue why 
taking a vacuum state at the beginning of inflation to evaluate this 
correlator seems a good approximation. 

In section \ref{slowroll} we analyze the behaviour of the background scalar 
fields during slow-roll inflation in various multiple field cases.  In
subsection \ref{SlowRollFlat} two examples of a quadratic  potential on a flat
manifold are considered: with equal and with different  masses.  Next, we turn
to the generalized situation of a curved manifold  with arbitrary potential in
subsection \ref{SlowRollCurved}.  To illustrate some aspects we take  the
curved manifold to be the sphere with embedding coordinates.  In each of these
examples we compute the model dependent factor  that appears in the expression
for the correlation function of the  gravitational potential. 

\np

\section{Scalar perturbations}
\labl{scalpert}

\subsection{Linearized gravitational perturbations}

We can divide general perturbations in the universe in the following two types:
metric perturbations and matter perturbations. Of course they are related by
the Einstein equations. On the other hand we can divide both matter and metric
perturbations in three different classes: scalar, vector, and tensor
perturbations \cite{Bardeen,Stewart,Mukhanovetal}, depending on how they
transform under spatial transformations of the background metric. In this paper
we consider only scalar perturbations since
they are the main cause of the fluctuations in the CMBR. We assume throughout
this paper that all perturbations are small, as on the one hand they presumably
originate from quantum perturbations, while on the other hand the fluctuations
in the CMBR that we observe are tiny. In particular this means that we 
linearize all equations with respect to the perturbations. This
section is essentially a review of \cite{Mukhanovetal}.

The Robertson-Walker metric for a spatially flat background combined 
with scalar metric perturbations may be written as 
\equ{
g_{\mu\nu}^{\mathrm{full}}(\eta,\vc{x}) 
= g_{\mu\nu}(\eta) + \gd g_{\mu\nu}(\eta,\vc{x}) =
a^2 \pmtrx{-1 & 0\\ 0 & \gd_{ij}}
+ a^2 \pmtrx{-2 \Phi & B_{,j}\\ B_{,i} & -2 \Psi \gd_{ij} + 2 E_{,ij}}.
}
We take a flat background with scale factor $a(\eta)$, since we are going to 
apply our formulae to the universe during and after inflation, when it has
already been inflated to complete flatness.
$\Phi,\Psi,B,E$ are four scalar functions of spatial coordinates $\vc{x}$ 
and conformal time $\eta$ which together describe the metric perturbations.
One often refers to $\gF$ as the Newtonian potential, as the 
$00$-component of the metric in a weak field approximation can be 
identified with the potential of Newtonian gravity.
The conformal time $\eta$ is related to the comoving time $t$ by 
$\d t = a \d \eta$. The advantage of using conformal time is that then 
the scale factor $a$ is an overall factor of the full metric, not just of the 
spatial part. We take the scale factor to have dimension of length and 
therefore $\get$ and $\vc x$ are dimensionless. 
Differentiation with respect to conformal and comoving 
time are denoted by $^\prime \equiv \der_\get$ and $\dot{} \equiv \der_t$, 
respectively.  
The conformal Hubble parameter $\cH$ is defined as $\cH \equiv a'/a$. 
It is related to the comoving Hubble parameter $H \equiv \dot{a}/a$ by 
$\cH = a H$. The notation~$_{,i}$ denotes a derivative 
with respect to the spatial coordinate $x^i$. 

There is a problem with the interpretation of the metric perturbations because 
it is difficult to separate the physical metric perturbations from the ones
that can be gauged away by a coordinate transformation. Of course the final, 
physical results do not depend on the choice of coordinates.  However, it is
often convenient if intermediate results are also independent of the
gauge chosen. To this end so-called gauge-invariant quantities are introduced
\cite{Bardeen,Mukhanovetal}, which are defined to be gauge invariant with
respect to infinitesimal coordinate  redefinitions. In this approach one
defines  gauge-invariant metric and matter quantities and uses those in the
full system of metric and matter perturbations.\footnote{The main alternative
is the fluid flow approach  \cite{LiddleLyth,LythRiotto}, where one makes
certain  assumptions about the matter content of the universe, which enables
one to eliminate the metric perturbations and derive equations for the matter
perturbations in closed form.}  They are:
\equ{
\Phi^{(gi)} = \Phi + \frac{1}{a} \left [ (B-E') a \right ]', \quad
\Psi^{(gi)} = \Psi - \cH (B-E'), \quad
\gd q^{(gi)} = \gd q + q' (B-E').
\labl{deltaphigi}
}
Here we have separated an arbitrary scalar (matter) quantity
$q_{\mathrm{full}}(\eta,\vc{x})$ into 
a homogeneous background part $q(\eta)$
and a perturbation $\gd q(\get, \vc{x}) = q_{\mathrm{full}}(\get, \vc{x}) 
- q(\get)$, just as we did for the metric.
%One should note that these so-called gauge-invariant quantities are only
%invariant under infinitesimal scalar coordinate transformations
%\equ{
%\eta \ra \eta + \xi^0(\eta,\vc{x}), \qquad\qquad
%x^i \ra x^i + \gd^{ij} \xi_{,j}(\eta,\vc{x}).
%}
%Also, the choice made in (\ref{deltaphigi}) is not unique: $\gF^{(gi)}$ and 
%$\gPs^{(gi)}$ are independent, so that any other gauge invariant function of 
%the metric perturbation is written as a function of those two.
As can be seen from (\ref{deltaphigi}), working with these gauge-invariant
quantities is equivalent to choosing the longitudinal gauge in which $B=E=0$. 
As we only employ the longitudinal gauge in this paper, we 
omit the $(gi)$ labels without risk of confusion. 

For most of the treatment in this paper it is irrelevant whether  scalar
perturbations like $\gd q(\get, \vc x)$ and $\Phi(\eta,\vc{x})$ are classical
or quantum objects. This is because we will linearize in those  quantities so
that the quantum nature (such as variables that do not commute) does not  play
a role. Hence we may  derive and manipulate the equations as if all quantities
were classical.  Only when we are computing the correlator of the Newtonian
potential  $\langle \gF \gF \rangle$ in section~\ref{spectrum} do we have to
take the quantum nature of the perturbations into account.

We now treat the dynamics of the universe with scalar perturbations. 
The background Einstein equations read
\equ{
\cH^2  =  - \frac{1}{3} \gk^2 a^2 T^0_{\;0},
\qquad\qquad
\lh 2 \cH' + \cH^2 \rh \gd^i_j  =  - \gk^2 a^2 T^i_{\;j},
\labl{EinstBack}
}
and $T^0_{\;i}=0$.
Expanding the Einstein equations to first order in the perturbations gives
\equ{
\equa{
-3 \cH(\cH \Phi + \Psi') + \gD \Psi & \ = \
- \half \gk^2 a^2 \gd T^0_{\;0},
\labl{Einstein00}\\
\lh \cH \Phi + \Psi' \rh_{,i} & \ = \ 
- \half \gk^2 a^2 \gd T^0_{\;i},
\labl{Einstein0i}
}\\
\!\!\!
\left [ (2 \cH' + \cH^2)\Phi + \cH \Phi' 
+ \Psi'' + 2 \cH \Psi' + \half \gD (\Phi - \Psi) \right ] \gd^i_j
- \half \gd^{ik} (\Phi - \Psi)_{,kj} \ = \
\half \gk^2 a^2 \gd T^i_{\;j},
\labl{Einsteinij}
}
with $\gk^2 \equiv 8\pi G = 8\pi/M_P^2$ and $\gD = \sum_i \der_i \der_i$.
Later on we often switch to complex Fourier modes $f_{\vc k}(\get)$, defined by
\equ{
f(\eta,\vc{x}) = \frac{1}{(2\pi)^{3/2}} \int \d^3 \vc{k} \lh f_\vc{k}(\eta)
e^{- i \vc{k} \vc{x}} + f_\vc{k}^*(\eta) e^{i \vc{k} \vc{x}} \rh,
\labl{Fourier}
}
where $f$ is any real quantity that depends on both time and space coordinates,
e.g.\ $\Phi(\eta,\vc{x})$. 
After this switch equations for $f(\eta,\vc{x})$
become equations for $f_\vc{k}(\eta)$ and the spatial Laplacian $-\gD$ is 
replaced by $k^2 = |\vc{k}|^2$. 
%We often 
%omit the subscript $_\vc{k}$, but it should be clear from the context whether 
%we mean Fourier modes or not.

The complicated system of Einstein equations is simplified considerably 
when the matter is described by a scalar field theory.  
For a scalar field theory with an arbitrary number of fields one
can easily verify that $\gd T^i_{\;j} \propto \gd^i_j$ to first order in the 
perturbations, while $\gd T^0_{\;i}$ can be written as 
$\gd T^0_{\;i} = \gd F_{,i}$ to the same order.
Here $\gd F$ is a scalar function of the fields and their perturbations which 
depends on the scalar field theory.
A very important simplification then follows from
an argument by Mukhanov et al.\ \cite{Mukhanovetal}, who show that one can take
$\Psi=\Phi$ if $T^i_{\;j} \propto \gd^i_j$. 

By taking a normalized trace of the $(ij)$-components
and subtracting the $(00)$-component of the background Einstein 
equations \eqref{EinstBack}, one obtains the following equation:
\equ{
\cH^2 -\cH' = \half \gk^2 a^2 \lh \frac{1}{3} T^i_{\;i} - T^0_{\;0} \rh.
\labl{Einstback}
}
A similar procedure with the perturbed components of the Einstein equations 
(\ref{Einstein00}) and (\ref{Einsteinij}), setting 
$\gPs = \gF$, gives the equation of motion for the Newtonian potential $\gF$
\equ{
\Phi'' + 6 \cH \Phi' + 2 (\cH' + 2\cH^2) \Phi - \gD \Phi
= \half \gk^2 a^2 \lh \frac{1}{3} \gd T^i_{\;i} + \gd T^0_{\;0} \rh.
\labl{eqPhi}
}
With a suitable choice for the Newtonian potential $\gF$ to eliminate the
integration constant the perturbed
$(0i)$-component \eqref{Einstein0i} of the Einstein equations can be 
integrated using $\gd T^0_{\;i} = \gd F_{,i}$:
\equ{
\gF' + \cH \gF = - \frac 12 \gk^2 a^2 \gd F.
\labl{EinstInt0i}
}
In the next subsection we apply these equations to the case of a 
scalar field theory with a single field, and in section \ref{sep_multi_grav} 
to the multiple field case.

\subsection{Scalar perturbations due to a single scalar field}

Now we consider the case of a single real scalar field, both to review
the method of Mukhanov et al.\ \cite{Mukhanovetal} and to be able to compare it 
with the multiple scalar field case when we treat the latter in 
section~\ref{multi}.
From the Lagrangean 
\equ{
\cL = \sqrt{-g} \lh - \half \der^\mu \phi \der_\mu \phi - V(\phi) \rh
}
we obtain the equation of motion for the scalar field,
\equ{
D^\gm \der_\gm \gf - V_{,\gf} = 0, 
\labl{eqmot1}
}
where $D_\gm$ is the covariant space-time derivative, 
and the energy-momentum tensor
\equ{
T^\mu_{\;\nu} = \der^\mu \phi \der_\nu \phi - \gd^\mu_\nu \lh \half \der^\gl
\phi \der_\gl \phi + V(\phi) \rh.
}
In these three equations, $\gf$ denotes the total field $\gf_{\mathrm{full}}$.
Now we separate the background from the perturbations, as defined
below (\ref{deltaphigi}), so that in the remaining equations $\gf$ denotes the
background part of the field.
The integrated $(0i)$-Einstein equation \eqref{EinstInt0i},
\equ{
\Phi' + \cH \Phi = \half \gk^2 \phi' \gd\phi,
\labl{Einstein0i1}
}
and the background equation of motion for the scalar field,
\equ{
\phi'' + 2 \cH \phi' + a^2 V_{,\phi} = 0,
\labl{eqmotback1}
}
can be used to eliminate the fluctuation $\gd \gf$ and 
the first derivative of the potential $V_{,\gf}$ from
the equation of motion \eqref{eqPhi} for the Newtonian potential $\gF$,
\equ{
\Phi'' + 6 \cH \Phi' + 2 (\cH' + 2\cH^2) \Phi - \gD \Phi
= - \gk^2 a^2 V_{,\phi} \gd\phi. 
\labl{eqPhi1}
}
Then this last equation takes the form of a homogeneous differential
equation:
\equ{
\Phi'' + 2 \lh \cH - \frac{\phi''}{\phi'} \rh \Phi' 
+ 2 \lh \cH' - \cH \, \frac{\phi''}{\phi'} \rh \Phi - \gD \Phi = 0.
\labl{eqPhi1a}
}
Using equation \eqref{Einstback},
\equ{
\cH^2 -\cH' = \half \gk^2 {\phi'}^2,
\labl{Einstback1}
}
equation \eqref{eqPhi1a} can be written in terms of 
$u \equiv \frac {a}{\gk^3 \phi'}\Phi$  as
\equ{
u'' - \frac{\gth''}{\gth} \, u - \gD u = 0,
\qquad\text{with}\qquad
\gth \equiv \frac {\cH}{a\phi'}.
\labl{equ1}
}
The factor $\gk^{-3}$ in the definition of $u$ has been 
introduced to give it mass dimension one. 
We can use (\ref{Einstback1}) and the relation $\cH=aH$ to obtain expressions 
for $u = \Phi/(\gk^2 \sqrt{-2\dot{H}})$ and $\gth= \gk H/(a\sqrt{-2\dot{H}})$ 
that do not contain the scalar field $\gf$, but only quantities that are
well-defined also after inflation. 
Equation \eqref{equ1} for the variable $u$ is very important because
it can be used throughout the evolution of the universe: during scalar
field, radiation, and matter domination \cite{Mukhanovetal} (only during
matter domination an extra factor containing the sound velocity has to be
added). 

By varying (\ref{eqmot1}) to first order in the perturbations we find the
equation of motion for the scalar field fluctuations:
\equ{
\lh \der_\eta^2 + 2 \cH \der_\eta - \gD + a^2 V_{,\phi\phi} \rh \gd\phi
= 4 \phi' \Phi' - 2 a^2 V_{,\phi} \Phi.
\labl{eqmotpert1}
}
Notice that we did not need to use this equation in our derivation of a 
homogeneous equation for $\gF$ \eqref{eqPhi1a} or $u$ (\ref{equ1}). 
It was not needed, because the equation of motion of the scalar field can be
derived from the constraint that the energy-momentum 
tensor is divergenceless, $D_\mu T^{\mu\nu} = 0$, and is therefore not an
independent equation. 
This is closely related to the fact that we could solve for $\gd \gf$ 
by dividing the integrated $(0i)$-Einstein equation \eqref{Einstein0i1} 
by the velocity $\gf'$. 
In the case of more fields this constraint can no longer 
reproduce all equations of motion, so we expect to need the equations of
motion for the field perturbations in that case.

\np

\section{Inflation with multiple scalar fields}
\labl{multi}

\subsection{Geometrical concepts}
\labl{intermezzo}

We now turn to the multiple scalar field case, where the scalars 
$\Bgf = (\gf^a)$ can be interpreted as the coordinates of a real manifold 
$\cM$ on which a metric $\mx{G}$ is defined. 
To make optimal use of the geometrical structure of this manifold 
in our discussion of the dynamics of scalar fields and 
their perturbations, we need to introduce some geometrical concepts.
From the components of metric $G_{ab}$ 
the metric-connection $\gG^a_{bc}$ is obtained using the metric postulate. 
The definition of the manifold $\cM$ is coordinate 
independent, therefore the description of this manifold is 
invariant under non-singular local coordinate transformations
\equ{
\gf^a \lra \tgf^a = X^a(\gf).
\labl{transf}
}
In our treatment of scalar perturbations due to multiple scalar fields 
we heavily rely on the concept of tangent vectors. 
A vector $\vc{A} = (A^a)$ is called a vector in the 
tangent space $T_p\cM$ at a point $p \in \cM$ if it 
transforms as
\equ{
A^a \lra \tA^a = X^a_{\; b}(\gf) A^b, 
\qquad\qquad
X^a_{\; b}(\gf) = X^a_{\; ,b}(\gf),
}
where the comma denotes differentiation with respect to
local coordinates. A simple example of a (tangent) vector is the 
differential $\text{d} \Bgf$.
The cotangent space is the dual of the tangent space. 
Its elements are linear operators on the tangent space 
\equ{
\mtrx{
^\ast\vc{C}: & T_p\cM  & \lra    & \Real \\
         & \vc{A} & \mapsto & C_a A^a
}
}
As $C_a A^a$ is a scalar object, the cotangent vector
$^\ast\vc{C}$ transforms as
\equ{
C_a \lra \tC_a = C_b (X\inv)^b_{\; a}.
}
The metric $G_{ab}$ can be used to construct a cotangent 
vector
\(
(\vc A^\dag)_a \equiv A^b G_{ba}
\)
from the tangent vector $\vc{A}$. Using index-free 
notation this reads  $\vc A^\dag = \vc A^T \mx G$. 
The notion of (co)tangent vectors defined at a point 
$p \in \cM$ can be extended over the whole manifold $\cM$ by 
interpreting them as sections of the (co)tangent bundle.

Using the metric $\mx{G}$ we introduce an inner product of 
two vectors $\vc A$ and $\vc B$ on the tangent space of the manifold 
and the corresponding norm 
\equ{
\vc{A} \cdot \vc{B} \equiv 
\vc A^\dag \vc B = 
\vc{A}^T \mx{G} \vc{B} = A^a G_{ab} B^b, 
\qquad\qquad
|\vc{A}| \equiv \sqrt{\vc{A}\cdot\vc{A}}.
}
The Hermitean conjugate $\mx{L}^\dag$ of a linear operator 
$\mx{L}: T_p \cM \lra T_p \cM$ 
with respect to this inner product is defined by
\equ{
\vc{B} \cdot (\mx{L}^\dag \vc{A}) \equiv
(\mx{L} \vc{B}) \cdot \vc{A},
}
so that $\mx{L}^\dag = \mx{G}^{-1} \mx{L}^T \mx{G}$.
A Hermitean operator $\mx H$ satisfies 
$\mx{H}^\dag = \mx{H}$. 
An important example of Hermitean operators are the projection 
operators. Apart from being Hermitean, a projection operator $\mx P$ 
is idempotent: $\mx{P}^2 = \mx{P}$.

To complete our discussion on the geometry of $\cM$ we
introduce different types of derivatives.
In the first place we have the covariant derivative on the manifold, 
denoted by $_{;a}$ or $\nabla_a$, which acts in the usual way, i.e.\ 
\equ{
A^a_{\; ;b} = \nabla_b A^a \equiv A^a_{\; ,b} + \gG^a_{bc} A^c
}
on a vector $A^a$, and $V_{;a} = \nabla_a V = V_{,a}$ on a scalar $V$.
It is convenient to also introduce index-free notation for 
(covariant) derivatives. 
On a scalar function $V$ (e.g.\ the potential), the 
derivative $\Bder$ and the covariant derivative $\Bnabla$ are 
equal
\equ{
(\Bnabla V)_a = (\Bder V)_a \equiv V_{,a}.
}
Since we represent $\text{d} \Bgf$ as a standing vector, $\Bnabla$ and 
$\Bder$ are naturally lying vectors and therefore $\Bnabla^T$ and 
$\Bder^T$ are standing vectors. The second covariant derivative of a 
scalar function $V$ is a matrix with two lower indices:
\equ{
(\Bnabla^T \Bnabla V)_{ab} = \nabla_a \nabla_b V.
}
Of course, the same holds good for ordinary derivatives $\Bder$. 

The curvature tensor of the manifold can be introduced 
using tangent vectors $\vc{B}, \vc{C}, \vc{D}$:
\equ{
[\mx{R}(\vc{B},\vc{C})\vc{D}]^a \equiv
%\Bnabla_{\vc{C}}\Bnabla_{\vc{D}} \vc{B} -
%\Bnabla_{\vc{D}}\Bnabla_{\vc{C}} \vc{B} -
%\Bnabla_{[ \vc{C}, \vc{D}]} \vc{B} \non\\
%\Longleftrightarrow \quad 
R^a_{\; bcd} \, B^b C^c \, D^d \equiv
\lh \gG^a_{bd,c} - \gG^a_{bc,d} + \gG^e_{bd} \gG^a_{ce}
- \gG^e_{bc} \gG^a_{de} \rh  B^bC^c \, D^d.
}
%where $\Bnabla_{\vc{B}}$ is the derivative in the direction
%$\vc{B}$: $\Bnabla_{\vc B} = B^a \nabla_a$. 
One should realize that for later notational convenience we do not use the 
standard definition as made for example in \cite{Nakahara}: 
our $\mx{R}(\vc{B},\vc{C})\vc{D}$ is conventionally 
denoted by $\vc{R}(\vc{C},\vc{D},\vc{B})$.

Next we discuss how spacetime derivatives act on spacetime dependent 
tangent vectors and their derivatives.
Purely spacetime covariant derivatives are denoted by $D_\gm$ and 
are defined in the usual way.
The covariant derivative $\cD_\gm$ on a vector $\vc{A}$ of 
the tangent bundle is defined in components as
\equ{
\cD_\mu A^a \equiv
\der_\gm A^a + \gG^a_{bc} \der_\gm \gf^b A^c,
}
while $\cD_\gm$ acting on a scalar is simply equal to $\der_\gm$.

After the introduction of this standard geometrical machinery, we  now develop
some concepts to describe a time dependent scalar field  background. They are
used when we consider multiple   field inflation. 
Consider a curve $\Bgf(t)$ on a manifold $\cM$,
parameterized by  a real variable $t$. In later sections, when we describe the 
time evolution of coupled systems consisting of multiple scalar fields and
gravity, this variable $t$ is interpreted as comoving time.  Along
this curve the $n$th derivative vector can be defined by 
\equ{ 
\Bgf^{(1)} \equiv \Dot\Bgf = \frac {\text{d}\Bgf}{\text{d} t}
\qquad\mbox{and}\qquad  
\Bgf^{(n)} \equiv \cD_t^{(n-1)} \Dot \Bgf   \quad \text{for} \quad  n \geq 2. 
} 
In applications in later sections $\Bgf^{(1)}$ and $\Bgf^{(2)}$ represent the
velocity and acceleration of the background scalar fields, respectively. 
In general the  vectors $\Bgf^{(1)},
\Bgf^{(2)}, \ldots$ do not point in the same  direction. From these vectors a
set of  orthonormal unit vectors  is obtained by  using the Gram-Schmidt
orthogonalization process. The first unit  vector $\vc{e}_1$ is given by the
direction of $\Bgf^{(1)}$. The  second unit vector $\vc{e}_2$ is determined by
that part of  $\Bgf^{(2)}$ that is perpendicular to $\vc{e}_1$, and so on. To
obtain the  direction of $\Bgf^{(2)}$ perpendicular to $\vc{e}_1$, we use the
projection  operators $\mx{P}_1$ and $\mx{P}_1^\perp$ that project on subspaces
parallel and perpendicular to $\Bgf^{(1)}$, respectively, and require  that
$\vc{e}_2$ is proportional to $\mx{P}_1^\perp \Bgf^{(2)}$.  

These definitions can be extended to $\vc{e}_n, \mx{P}_n$, etc., for 
any $n$. 
The unit vector $\vc{e}_n$ points in the direction of $\Bgf^{(n)}$ 
perpendicular to the first $n-1$ unit vectors 
$\vc{e}_1,\ldots, \vc{e}_{n-1}$.
The operator $\mx{P}_n$ projects on $\vc{e}_n$ and  
$\mx{P}_n^\perp$ projects on the subspace  which is perpendicular to 
$\vc{e}_1, \ldots, \vc{e}_n$. This subspace is also perpendicular to 
the derivative vectors $\Bgf^{(1)}, \ldots, \Bgf^{(n)}$. 
To obtain all these objects at once, we let $\mx{P}_{0}^\perp = \Id$ 
be the identity and define the mutually orthonormal unit vectors 
$\vc{e}_n$ (for $n=1,2,\ldots$) from $\Bgf^{(n)}$ and the projection operator 
$\mx{P}_{n-1}^\perp$ by
\equ{
\gf^{(n)}_{~n} =| \mx{P}_{n-1}^\perp \Bgf^{(n)} |,
\qquad
\vc{e}_n =
\frac{ \mx{P}_{n-1}^\perp \Bgf^{(n)} }{ \gf^{(n)}_{~n} },
\qquad 
\mx{P}_n = \vc{e}^{\;}_n \vc{e}_n^\dag, 
\qquad
\mx{P}_n^\perp = \Id - \sum_{q=1}^n \mx{P}_q.
}
By construction the vector $\Bgf^{(n)}$ can be expanded 
in these unit vectors as 
\equ{
\Bgf^{(n)} = 
\left( 
\mx{P}_1 + \ldots + \mx{P}_{n} 
\right) \Bgf^{(n)} = 
\sum_{p = 1}^n \gf^{(n)}_{~p} \vc{e}_p,
\qquad
\gf^{(n)}_{~p} = \vc{e}_p \cdot \Bgf^{(n)}.
}
In particular, we have that 
\(
\gf^{(n)}_{~n} = \vc{e}_n \cdot \Bgf^{(n)} = 
| \mx{P}_{n-1}^\perp \Bgf^{(n)} |.
\)
As the projection operators $\mx{P}_1$ and $\mx{P}_1^\perp$ will 
occur frequently in later sections, we introduce the short-hand notation:
\equ{
\mx{P}^\parallel = \mx{P}_1 = \frac{\Dot{\Bgf} \, \Dot{\Bgf}^\dag}
{|\Dot{\Bgf}|^2}, \qquad\qquad
\mx{P}^\perp = \mx{P}_1^\perp = \Id - \mx{P}^\parallel.
\labl{projectors}
}
In terms of these two operators we can write a general vector and matrix as
follows: 
\equ{ 
\vc{A} = \vc{A}^\parallel + \vc{A}^\perp, 
\qquad\qquad
\mx{M}  =  \mx{M}^{\parallel\parallel} + \mx{M}^{\parallel\perp} +  
\mx{M}^{\perp\parallel} + \mx{M}^{\perp\perp},
\labl{vecmatproj} 
} 
with $\vc A^\parallel = \mx P^\parallel \vc A$ and 
$\mx{M}^{\parallel\,\parallel} \equiv 
\mx{P}^\parallel \mx{M} \mx{P}^\parallel$, etc. 
Notice that because of the hermiticity of the projection operator 
$\vc{A}^\perp \cdot \vc{B} = \vc{A} \cdot \vc{B}^\perp = \vc{A}^\perp 
\cdot \vc{B}^\perp$.

\subsection{Multiple scalar fields and gravitational perturbations}
\labl{sep_multi_grav}

We now consider scalar fields $\Bgf$ that are the local coordinates 
of a manifold $\cM$, using the geometrical concepts introduced 
in the previous section. The space-time derivative of the background field
$\der_\mu \Bgf$ transforms as a vector, even though the
fields $\Bgf$ in general do not, as they are coordinates on a manifold.
Also the field perturbation $\Bgd\Bgf$ and its gauge-invariant form defined
by (\ref{deltaphigi}) transform as vectors.
Only the covariance with respect to the coordinate transformations
(\ref{transf}) of the target space $\cM$ is manifest in our treatment,
because we use a flat Robertson-Walker background and work to first order in
the perturbations.

The Lagrangean for the scalar field theory with potential $V$ on the 
manifold $\cM$ can be written as
\equ{
\cL_\cM = 
\sqrt{-g} \lh - \half \der^\gm \Bgf \cdot \der_\gm\Bgf - V \rh
%=
%\sqrt{-g} \lh - \half G_{ab} g^{\mu\nu} \der_\mu \phi^a \der_\nu \phi^b
%- V(\phi) \rh
.
\labl{Lagrangean}
}
The equations of motion for the scalars are given by
\equ{
g^{\mu\nu} \lh \cD_\mu \gd^\gl_\nu - \gG^\gl_{\mu\nu} \rh \der_\gl \Bgf 
- \mx G\inv \Bnabla^T V = 0,
} 
and the energy-momentum tensor is
\equ{
T^\mu_{\;\nu} = \der^\mu \Bgf \cdot \der_\nu \Bgf
- \gd^\mu_\nu \lh \half \der^\gl \Bgf \cdot \der_\gl \Bgf + V \rh.
\labl{Tmunu2}
}
In these three equations, $\Bgf$ denotes the total field $\Bgf_{\mathrm{full}}$.
Now we separate the background from the perturbations, as defined
below (\ref{deltaphigi}), so that from now on $\Bgf$ will always denote the
background part of the field.
In this case of multiple scalar fields the equation of motion 
\eqref{eqPhi} of the Newtonian potential $\gF$ reads to first order
in the perturbations 
\equ{
\Phi'' + 6 \cH \Phi' + 2 (\cH' + 2\cH^2) \Phi - \gD \Phi
= - \gk^2 a^2 (\Bnabla V \, \Bgd\Bgf). 
\labl{eqPhi2}
}
The integrated $(0i)$-component \eqref{EinstInt0i} 
of the Einstein equations takes the form
\equ{
\Phi' + \cH \Phi = \half \gk^2 \Bgf' \cdot \Bgd\Bgf.
\labl{Einstein0i2}
}

In this case it is not possible to construct the analogue of  the single field
homogeneous equation of motion \eqref{eqPhi1a} for the Newtonian potential
$\gF$, because (\ref{Einstein0i2})  no longer  contains a simple multiplication
of two scalars,  but an inner product of two  vectors.  Therefore, one cannot
extract an explicit expression for $\Bgd\Bgf$, as was the case in equation
\eqref{Einstein0i1} for the single field situation. To overcome this
difficulty, we divide the field perturbation $\Bgd\Bgf$ in a part that is
parallel to the velocity field $\Bgf'$ and a part that is perpendicular, using
the projection operators (\ref{projectors}) defined in the previous
section.\footnote{A similar decomposition in the case of two field inflation was
simultaneously developed in \cite{Gordonetal}.} 
Here we use that for the projection operators $\mx{P}^\parallel$ and
$\mx{P}^\perp$ there is no difference between using comoving time or
conformal time: they only depend on the direction of $\Dot\Bgf$, which is the
same  as the direction of $\Bgf'$.  Once we have separated the fields in this
way, the parallel part  $\Bgd\Bgf^\parallel$ can be eliminated in a way 
analogous to the single field case.

Using  the integrated $(0i)$-component of the Einstein equations 
\eqref{Einstein0i2} together with the background equation of motion for the
scalar fields,
\equ{
\cD_\eta \Bgf' + 2 \cH \Bgf' + a^2 \mx G\inv \Bnabla^T V = 0,
\labl{eqmotback2}
}
the right-hand side of equation \eqref{eqPhi2} for $\gF$ can be rewritten as
\equa{
- \gk^2 a^2 (\Bnabla V \, \Bgd\Bgf)  & \ = \  
- \gk^2 a^2 (\mx{G}\inv \Bnabla^T V) \cdot \Bgd\Bgf \ = \
\gk^2 \lh \cD_\eta \Bgf' + 2 \cH \Bgf' \rh \cdot 
\lh \vc{e}_1 (\vc{e}_1 \cdot \Bgd\Bgf) + \Bgd\Bgf^\perp \rh 
\non \\
& \ = \  2 (\Phi' + \cH \Phi) \lh \frac{1}{|\Bgf'|} (\cD_\eta \Bgf')
\cdot \vc{e}_1 + 2 \cH \rh
+ \gk^2 (\cD_\eta \Bgf') \cdot \Bgd\Bgf^\perp,
}
where we used the definition of the projection operators (\ref{projectors}).
Inserting this expression in \eqref{eqPhi2} and realizing that 
\(
|\Bgf'|' \, |\Bgf'| = (\cD_\eta \Bgf') \cdot \Bgf',
\)
we get
\equ{
\Phi'' + 2 \lh \cH - \frac{|\Bgf'|'}{|\Bgf'|} \rh \Phi'
+ 2 \lh \cH' - \cH \frac{|\Bgf'|'}{|\Bgf'|} \rh \Phi - \gD \Phi
= \gk^2 (\cD_\eta \Bgf') \cdot \Bgd\Bgf^\perp.
\labl{eqPhi2a}
}
Notice that, apart from the right-hand side, equation (\ref{eqPhi2a})
looks identical to equation (\ref{eqPhi1a})
for the single field case. However, it is exactly this right-hand side which
makes it necessary to solve a coupled system of differential equations, as
opposed to the decoupled system in the single field case. In this case we
need the equation of motion for the scalar field fluctuations,
\equ{
\lh \cD_\eta^2 + 2 \cH \cD_\eta - \gD - \mx{R}(\Bgf',\Bgf') 
+ a^2 \mx{M}^2(\Bgf) \rh \Bgd\Bgf
=  4 \Phi' \Bgf' - 2 a^2 \Phi\, \mx G\inv \Bnabla^T V,
\labl{eqmotpert2}
}
where we have introduced the mass-matrix
\equ{
\mx M^2 \equiv \mx G\inv \Bnabla^T \Bnabla V, 
\qquad\qquad
(\mx{M}^2)^a_{\; b}(\Bgf) = G^{ac}(\Bgf) V_{;cb}(\Bgf).
\labl{massmatrix}
}
Notice that $\mx{R}(\Bgf',\Bgf') = a^2 \mx{R}(\Dot{\Bgf},\Dot{\Bgf})$, so that
it is possible to absorb the curvature term into an effective mass matrix.
Equation \eqref{eqmotpert2} is the multiple field generalization of
(\ref{eqmotpert1}). However, we do not need the total perturbations, but only
the perpendicular part, as can be seen from (\ref{eqPhi2a}). This system of
equations for  $\Phi$ and $\Bgd\Bgf^\perp$ we will analyze in
section~\ref{Pertsolv}, but before we do that we consider the background
equations during  slow-roll inflation.

\subsection{Multiple field slow-roll functions}
\labl{slowrollsec}

Slow-roll inflation is driven by a flat scalar field potential that acts as an
effective cosmological constant because of the small slope. In the case of a
single scalar field, the notion of slow roll is  well-established (see e.g.\
\cite{LiddleLyth, LythRiotto, LiddleParsonsBarrow}). In this paper we 
generalize this concept to multiple scalar fields in a geometrical way.
Afterwards we discuss how our slow-roll functions are related to the well-known
single field  slow-roll parameters.

To define the slow-roll functions we use comoving time $t$, since then 
the background equation does not contain the rapidly changing scale factor $a$.
The background equation of motion (\ref{eqmotback2}), the Friedmann equation
(\ref{EinstBack}), and equation (\ref{Einstback}) read in comoving time
\equ{
\cD_t \Dot{\Bgf} + 3 H \Dot{\Bgf} + \mx G\inv \Bnabla^T V = 0,
\qquad
H^2 = \frac{1}{3} \gk^2 \lh \half |\Dot{\Bgf}|^2 + V \rh,
\qquad
\Dot{H} = - \frac{1}{2} \gk^2 |\Dot{\Bgf}|^2.
\labl{eqmotbackFriedmann}
}
The system is said to be in the slow-roll regime if 
$|\cD_t \Dot{\Bgf}| \ll |3 H \Dot{\Bgf}|$ and $\half |\Dot{\Bgf}|^2 \ll V$.
A more precise definition is given below \eqref{eqmotbackiter}. 
Using (\ref{eqmotbackFriedmann}) the last condition can also be written as
$(-\Dot{H}) \ll \frac{3}{2} H^2$.

The functions
\equ{
\tge(\Bgf) \equiv - \frac{\Dot{H}}{H^2}, 
%= \frac{\frac{1}{2} \gk^2 |\Dot{\Bgf}|^2}{H^2},
\qquad\qquad
\tilde \Bget (\Bgf) \equiv \frac {\Bgf^{(2)}}{H |\Dot \Bgf| }, 
\qquad\mbox{and}\qquad
\tilde \Bgx (\Bgf) \equiv \frac {\Bgf^{(3)}}{H^2 | \Dot \Bgf| }
\labl{slowrollfun}
}
can be defined whether or not slow roll is valid. 
Since both $\tilde \Bget$ and $\tilde \Bgx$ are vectors, they can be 
decomposed in components using the unit vectors defined in 
section~\ref{intermezzo}. 
The components of these vectors in the directions 
$\vc{e}_1, \vc{e}_2$ and $\vc{e}_3$ are given by
\equa{
\tget^\parallel 
& = \vc{e}_1 \cdot \tilde \Bget = 
\frac{\cD_t \Dot{\Bgf} \cdot \Dot{\Bgf}} {H |\Dot{\Bgf}|^2}, 
%= \frac{|\Dot{\Bgf}|^{\boldsymbol{\cdot}}}{H |\Dot{\Bgf}|},
&
\tget^\perp
& = \vc{e}_2 \cdot \tilde \Bget = 
\frac{|(\cD_t \Dot{\Bgf})^\perp|}{H |\Dot{\Bgf}|}, 
\labl{slowrollfuncomp}
\\
\tgx^{\parallel}
& = \vc{e}_1 \cdot \tilde \Bgx = 
 \frac{\cD_t^2 \Dot{\Bgf} \cdot \Dot{\Bgf}}
{H^2 |\Dot{\Bgf}|^2},
&
\tgx_2 & = \vc{e}_2 \cdot \tilde \Bgx = 
\frac{\vc{e}_2 \cdot \cD_t^2 \Dot{\Bgf} }{H^2 |\Dot{\Bgf}|},
\qquad
\tgx_3  = \vc{e}_3 \cdot \tilde \Bgx = 
\frac{|\mx{P}^\perp_2(\cD_t^2 \Dot{\Bgf})|}{H^2 |\Dot{\Bgf}|}.
\non
}
Since $\tilde \Bgx$ in general has two directions perpendicular to $\vc{e}_1$, 
we cannot use the ambiguous notation $\tgx^\perp$. However, 
since $\tilde \Bgx$ is a vector, $\tilde{\Bgx}^\perp$ is defined. 
The components of $\tilde \Bgx$ are not needed for the
background equations, but will appear in the equations for the perturbations, 
see section \ref{Pertsolv}.
The functions $\tge$, $\tget^\perp$, and $\tgx_3$ are non-negative,
while $\tget^\parallel$, $\tgx^\parallel$, and $\tgx_2$ can also be negative 
(apart from the sign, $\tget^\parallel$ is equal to 
$\frac{|(\cD_t \Dot{\Bgf})^\parallel|}{H |\Dot{\Bgf}|}$).

In terms of these functions the Friedmann equation
(\ref{eqmotbackFriedmann}) reads
\equ{
H = \frac{\gk}{\sqrt{3}} \, \sqrt{V} \lh 1 - \frac{1}{3} \tge
\rh^{-1/2}.
\labl{Friedmann}
}
For a positive potential $V$ the function $\tge < 3$, as can be seen from its
definition. Inserting this expression and the functions (\ref{slowrollfun}) 
into the background equation gives
\equ{
\Dot{\Bgf} + \frac{2}{\sqrt{3}} \, 
\frac{1}{\gk} \mx G\inv \Bnabla^T \sqrt{V}
= - \sqrt{\frac{2}{3}} \, \sqrt{V} \, \frac{\sqrt{\tge}}{1-\frac{1}{3}\tge}
\lh \frac{1}{3} \tilde{\Bget} + \frac{\frac{1}{3} \tge \, \vc{e}_1}
{1+\sqrt{1-\frac{1}{3}\tge}} \rh.
\labl{eqmotbackiter}
}
From this equation and the one above, both of which are still exact, we can 
define precisely what is  meant by slow roll. Slow roll is valid if $\tge$, 
$\sqrt{\tge}\, \tget^\parallel$ and $\sqrt{\tge}\, \tget^\perp$ are (much) 
smaller than unity. For this reason $\tge$, $\tget^\parallel$ and $\tget^\perp$
are called slow-roll functions. However, $\tget^\parallel$ and $\tget^\perp$
could even be somewhat larger than one during slow roll, if $\tge$ is
sufficiently small. On the other hand, we will often use the somewhat stronger
condition that $\tge$, $\tget^\parallel$ and $\tget^\perp$ have to be small
individually.   The components of $\tilde \Bgx$  are called second order
slow-roll functions, and they are assumed to  be of an order
comparable to $\tge^2$, $\tge\tget^\parallel$, etc. If slow roll is valid, we
can expand in powers of these  slow-roll functions. For example, we can expand
the previous equation to lowest non-zero order in slow roll, which gives
\equ{
\Dot{\Bgf} + \frac{2}{\sqrt{3}} \, 
\frac{1}{\gk} \mx G\inv \Bnabla^T \sqrt{V}
= - \sqrt{\frac{2}{27}} \, \sqrt{V} \, \sqrt{\tge}
\left [
\half \tge \, \vc{e}_1 + \tilde \Bget
+ \cO(\tget^\parallel \tge, \tget^\perp \tge, \tge^{2}) 
\right ].
}

Next we derive some useful expressions for the slow-roll functions in terms of
conformal time. Using (\ref{slowrollfuncomp}) and (\ref{projectors}) we find
\equa{
& \frac{(\cD_\eta \Bgf')^\perp}{|\Bgf'|} 
%= \cH \tget^\perp \vc{e}_2 
= \cH \tilde{\Bget}^\perp,
&& \frac{(\cD_\eta^2 \Bgf')^\perp}{|\Bgf'|} 
% = \cH^2 ( 3 \tget^\perp \vc{e}_2 + \tgx_2 \vc{e}_2 + \tgx_3 \vc{e}_3 )
= \cH^2 ( 3 \tilde{\Bget} + \tilde{\Bgx} )^\perp,
\labl{slowrollrel} \\
%\frac{\cH'}{\cH^2} = 1 - \tge,
& \cH' = \cH^2 ( 1 - \tge),
\qquad
\frac{|\Bgf'|'}{|\Bgf'|} = \cH( 1 + \tget^\parallel),
&& 
\frac{|\Bgf'|''}{|\Bgf'|} = \cH^2
\lh 2 - \tge + 3 \tget^\parallel + (\tget^\perp)^2 + \tgx^\parallel \rh.
\non
}
Here we differentiated the expression
$|\Bgf'|' |\Bgf'| = \cD_\eta \Bgf' \cdot \Bgf'$ with respect to $\eta$ to find 
an expression for $|\Bgf'|''$.
Differentiating the slow-roll functions with respect to conformal time $\eta$ 
and using some of the above results we find
\equ{
\tge' = 2 \cH \tge ( \tge + \tget^\parallel ), 
\quad
(\tget^\parallel)' = \cH [ \tgx^\parallel 
+ (\tget^\perp)^2 + \tge \tget^\parallel - (\tget^\parallel)^2 ],
\quad
\cD_\get \tilde \Bget = \cH [ \tilde \Bgx + 
( \tge - \tget^\parallel) \tilde \Bget ].
\labl{dereps}
}
%Hence we could eliminate $\tget^\parallel$ from the
%equations by introducing $\tge'$ instead. 

The slow-roll functions (\ref{slowrollfun}) are all defined as  functions of
covariant derivatives of the velocity $\Dot\Bgf$, the velocity $\Dot{\Bgf}$
itself, and the Hubble parameter $H$.  If the zeroth order slow-roll
approximation works well,  that is if the right-hand side of
\eqref{eqmotbackiter} can be neglected, as  well as the $\tge$ in
(\ref{Friedmann}), then we can use these two equations to eliminate
$\Dot{\Bgf}$ and $H$ in favour of the potential $V$.  This is the way the
conventional single field slow-roll parameters are defined. However, this
conventional definition has the disadvantage that the slow-roll  conditions
become consistency checks. Hence, while we can expand the exact equations in
powers of the slow-roll functions, that is impossible by construction with the
conventional slow-roll parameters.\footnote{In the context of single field
inflation this fact was noted before and  discussed in detail in
\cite{LiddleParsonsBarrow}.} Therefore, we only show what our slow-roll
functions look like in terms of the potential in this approximation, but we do
not adopt it as the definition:
\equa{
\tge = \frac{1}{2 \gk^2} \frac{|\Bnabla V|^2}{V^2},
\qquad
\tget^\parallel - \tge & = 
- \frac{1}{\gk^2} \frac{\Bnabla V \, \mx{M}^2  \mx G\inv \Bnabla^T V}
{V |\Bnabla V|^2} = - \frac{1}{\gk^2}
\frac{\tr\left[ \lh\mx{M}^2\rh^{\parallel\parallel}\right]}{V}, 
\non \\
\tget^\perp & = 
\frac{1}{\gk^2} \frac{|\mx{P}^\perp \mx{M}^2 \mx G\inv \Bnabla^T V|}
{V \, |\Bnabla V|} = \frac{1}{\gk^2}
\frac{\sqrt{\tr\left[\lh\mx{M}^2\rh^{\parallel\perp} 
\lh\mx{M}^2\rh^{\perp\parallel}\right]}}{V}.
\labl{slowrollfun2}
}
The effective mass matrix $\mx{M}^2$ is defined in \eqref{massmatrix}. Here
$\mx{P}^\parallel$ projects along the direction determined by $\Bnabla V$, which
is to lowest order identical to the direction of $\Bgf'$.

In order to avoid confusion and for later comparison, we finish this section by
very explicitly  comparing the slow-roll functions we defined in
(\ref{slowrollfun}) with the  ones conventionally used in the single field
case, $\ge$ and $\get$:
\equ{
\ge = \frac{1}{2 \gk^2} \frac{V_{,\gf}^2}{V^2} = \tge, 
\qquad\qquad
\get = \frac{1}{\gk^2} \frac{V_{,\gf\gf}}{V} = - \tget^\parallel + \tge,
\labl{convsrpar}
}
where the last equalities in both equations are only valid
to lowest order in the slow-roll approximation. Of course, 
$\tget^\perp$ does not exist in the single field case, as there are no other
directions than the parallel one.

\subsection{Decoupling of the perturbation equations}
\labl{Pertsolv}

In this section we analyze the perturbation equations. Starting point are
equations (\ref{eqPhi2a}) for $\Phi$ and (\ref{eqmotpert2}) for $\Bgd\Bgf$.
First we rewrite these equations in such a way that we can draw some
important conclusions using the slow-roll functions defined in the previous 
section. It turns out that the equation for the (redefined) gravitational
potential decouples from the field perturbations up to first order in slow
roll. Next we concentrate on solving this equation.

We can write the system of the perturbation equations as a homogeneous
matrix equation for the vector $(\Phi,\Bgd\Bgf^\perp)$. However, to remove the
first order derivative term from the equation of motion for $\Phi$ we define new
variables:
\equ{
u \equiv \frac{a}{\gk^2 |\Bgf'|} \, \gk\inv \Phi = \frac{1}{\sqrt{2}} 
\frac{a}{\gk \cH \sqrt{\tge}} \, \gk\inv \Phi,
\qquad 
\Bgd\vc{v} \equiv \frac{a}{\gk^2 |\Bgf'|} \, \Bgd\Bgf^\perp = 
 \frac{1}{\sqrt{2}} \frac{a}{\gk \cH \sqrt{\tge}} 
\, \Bgd\Bgf^\perp.
\labl{newvars}
}
The additional factor of $\gk^{-1}$ in the definition of $u$ has been 
included to make the mass dimension of both $u$ and $\Bgd\vc{v}$ 
equal to one. Another important point to
note is that we chose our redefinitions in such a way that no relative
slow-roll factors have been introduced in the relation between $u$ and
$\Bgd\vc{v}$ as compared to the relation between $\Phi$ and $\Bgd\Bgf^\perp$.

We derive the $u$ component of the matrix equation here, as it is 
the component that we use most. The equation for
$\Bgd\vc{v}$ is more complicated, which is why we only give the result
here, and refer the reader to the appendix for the derivation.
We start with rewriting equation (\ref{eqPhi2a}) in terms of $u$ and obtain
\equ{
u'' - \gD u + \lh \frac{|\Bgf'|''}{|\Bgf'|} - 2 \lh \frac{|\Bgf'|'}{|\Bgf'|} 
\rh^2 -\cH^2 + \cH' \rh u 
= \gk |\Bgf'| \, \frac{(\cD_\eta \Bgf')^\perp}{|\Bgf'|} \cdot 
\frac{a}{\gk^2 |\Bgf'|}  \Bgd\Bgf^\perp.
\labl{equstep}
}
Using the conformal time version of the third equation in 
(\ref{eqmotbackFriedmann}),
\equ{
\cH^2 - \cH' = \half \gk^2 |\Bgf'|^2,
\labl{Einstback2}
}
and the definitions of $\Bgd\vc{v}$ and the slow-roll functions 
(\ref{slowrollrel}) we can rewrite this as
\equ{
u'' - \gD u + \cH^2 \lh -2\tge -\tget^\parallel
-2 (\tget^\parallel)^2 + (\tget^\perp)^2 + \tgx^\parallel \rh u
= \sqrt{2} \, \cH^2 \tge^\half \tilde \Bget^\perp \cdot \Bgd\vc{v}.
\labl{equ}
}
Combining this equation for $u$ with the result for $\Bgd\vc{v}$ from 
the appendix \eqref{eqapp} we get the following, still exact, matrix equation 
for the perturbations:
\equ{
\lh \boldsymbol{\mathfrak{D}_0} + \boldsymbol{\mathfrak{D}_1}
+ \boldsymbol{\mathfrak{D}_2} \rh \pmtrx{u \\ \Bgd\vc{v}} = \boldsymbol{0}.
\labl{matrixeq}
}
Here $\boldsymbol{\mathfrak{D}_0}$, $\boldsymbol{\mathfrak{D}_1}$, and 
$\boldsymbol{\mathfrak{D}_2}$ contain only terms up to order zero, one, and
two in the slow-roll functions, respectively. With our choice of the
slow-roll functions no higher orders occur. They are given by:
\equ{
\boldsymbol{\mathfrak{D}_0} \equiv \pmtrx{
\der_\eta^2 - \gD & 0\\
0 & \cD_\eta^2 + 2 \cH \cD_\eta - \gD - \mx{R}(\Bgf',\Bgf') 
+ a^2 (\mx{M}^2)^{\perp\perp}
},
}
\equ{
\boldsymbol{\mathfrak{D}_1} \equiv \pmtrx{
- \cH^2 ( 2 \tge + \tget^\parallel )
& 0\\
2\sqrt{2} \, \tge^{-1/2} \tilde \Bget^\perp \, \gD &
\lh \tget^\parallel \Id + \vc{e}_1 (\tilde{\Bget}^\perp)^\dag \rh 
\lh 2 \cH \cD_\eta + 3 \cH^2 \rh 
},
}
\equ{
\boldsymbol{\mathfrak{D}_2} \equiv \cH^2 \pmtrx{
- 2 (\tget^\parallel)^2 + (\tget^\perp)^2 + \tgx^\parallel &
-\sqrt{2} \, \tge^{1/2} (\tilde \Bget^\perp)^\dag \\
0 &
\lh (\tget^\perp)^2 + \tgx^\parallel \rh \Id 
+ 4 \tilde{\Bget}^\perp (\tilde{\Bget}^\perp)^\dag 
+ \vc{e}_1 (\tilde{\Bgx}^\perp)^\dag
}.
\labl{matrixlast}
}

From these equations one can draw the important conclusion that the 
redefined gravitational potential $u$ decouples from the perpendicular
components of the field $\Bgd\vc{v}$ up to and including first order in slow 
roll. The resulting equation for $u$ may be written in Fourier components
\eqref{Fourier} as 
\equ{
u_{\vc k}'' + \lh k^2 - \frac{\gth''}{\gth} \rh u_{\vc k} = 0,
\qquad\mbox{with}\qquad
\gth \equiv \frac{\cH}{a |\Bgf'|} = \frac{\gk}{\sqrt{2}} 
\frac{1}{a \sqrt{\tge}}.
\labl{equ2}
}
From the second expression for $\gth$ and (\ref{dereps}) it follows that
\equ{
\frac{\gth'}{\gth} = - \cH \lh 1 + \tge + \tget^\parallel \rh,
\qquad
\frac{\gth''}{\gth} = \cH^2 \lh 2\tge + \tget^\parallel + 2 (\tget^\parallel)^2
- (\tget^\perp)^2 - \tgx^\parallel \rh.
%= - \frac{1}{2} \frac{\tge''}{\tge} + \frac{3}{4} \lh
%\frac{\tge'}{\tge} \rh^2 + \cH \frac{\tge'}{\tge} + \cH^2 \tge.
\labl{gthdblgth}
}
We see that we got back equation (\ref{equ1}), which was derived in
\cite{Mukhanovetal}. However, here we have proved that this equation is not 
only valid in the single field case, but also in the multiple field case, up to
and including first order in slow roll, i.e.\ neglecting the term that is 
proportional to $\sqrt{\tge}\, \tget^\perp$. This is precisely one of the 
combinations of slow-roll functions that is small if slow roll is valid. 
Notice that even if slow roll is no longer valid, the specific combination
$\sqrt{\tge}\, \tget^\perp$ may still be small. This means that slow-roll
inflation ends because one of the other combinations of slow-roll functions
becomes large. We treat an example of this later on in section
\ref{flatmasses}: a quadratic potential where the mass difference between the
lightest mass and the other masses is large enough. On the other hand, we see
that the equation for the field perturbations  $\Bgd\vc{v}$ already depends on
the solution for $u$ at lower order, although compared with other terms this
coupling term will become smaller during  inflation because it does not contain
$\cH^2$. We defined  $u$ and $\Bgd\vc v$ in terms of $\Phi$ and
$\Bgd\Bgf^\perp$ in equation  \eqref{newvars} with the same powers of $\tge$,
so that the decoupling holds to the same order for $\Phi$ and $\Bgd\Bgf^\perp$.
In the literature, e.g.\ \cite{PolarskiStar,MukhanovSteinhardt,Gordonetal},
this leading order perturbation in $u$ is called the adiabatic mode, while the
perturbations associated with the perpendicular field components are related to
the so-called isocurvature modes. In this paper we restrict ourselves to
the adiabatic mode, but in a subsequent paper we plan to investigate the
perpendicular field equations and the isocurvature modes.

The decoupled equation for $u$ cannot be solved analytically 
in general. However, we can solve equation (\ref{equ2}) for $u$ 
exactly in the two limits of large and small $k$, as was observed 
in \cite{Mukhanovetal}. The two linearly independent solutions 
are denoted by $e_{\vc k}$ and its complex conjugate $e^*_{\vc{k}}$. In the 
small wavelength limit the solution is 
\equ{
e_{\vc{k}}(\eta) = e^{i k(\eta - \get_k)}
\qquad\qquad\qquad\qquad\text{for}\qquad 
{k}^2 \mg \Bigl | \frac{\gth''}{\gth} \Bigr |,
\labl{soluearly}
}
where the time $\get_k$ is defined as the time when $k^2$ is equal 
to $|\gth''_k/\gth_k|$. (We use the non-bold subscript $_k$ 
to indicate that a quantity is evaluated at $\eta=\eta_k$.) 
The normalization of the two independent solutions is such that 
their Wronskian satisfies
\equ{
W(e^*_{\vc{k}}(\get), e^{\;}_{\vc{k}}(\get)) = 
e^*_{\vc{k}}(\get) e^\prime_{\vc{k}}(\get) 
- e^{*\, \prime}_{\vc{k}}(\get) e^{\;}_{\vc{k}}(\get) = i\, 2 k. 
\labl{Wronskianupm}
}
In the limit of large wavelengths the solution of \eqref{equ2} is 
\equ{
e_{\vc{k}}(\eta) = C_{\vc{k}} \gth(\eta) 
+ D_{\vc{k}} \gth(\eta) 
\int_{\eta_k}^\eta \frac{\d\eta'}{\gth^2(\eta')}
\qquad\qquad\text{for}\qquad
{k}^2 \ml \Bigl | \frac{\gth''}{\gth} \Bigr |
\labl{solu},
}
which can be rewritten in terms of comoving time as
\equ{
e_{\vc{k}}(t) = \frac{H}{a |\dot{\Bgf}|} \lh C_{\vc{k}} - D_{\vc{k}} 
\frac{2 a_k}{\gk^2 H_k} \rh
+ \frac{2}{\gk^2 |\dot{\Bgf}|} \, D_{\vc{k}} 
\lh 1 - \frac{H}{a} \int_{t_k}^t a(\gt) \, \d \gt \rh.
\labl{longu}
}
Here we have used the definition of $\gth$, the relation
$\tge = (1/H)^{\boldsymbol{\cdot}}$, and integration by parts. 

Simply joining the two limits in a continuously differentiable 
way at $\get = \get_k$ determines the integration constants:
\equ{
C_{\vc{k}} = \frac{1}{\gth_k}
\qquad\text{and}\qquad
D_{\vc{k}} = 
\gth_k \left [ ik - \frac{\gth'_k}{\gth_k} \right ].
}
This joint solution gives a good approximation of the true solution of
\eqref{equ2} if slow roll is valid, because then the transition region is 
small, as we now show.  First, we define the transition region as that region
where the terms $k^2$ and $\gth''/\gth$ in the equation of motion for $u$ are
of the same order, i.e.\ within a factor $\ga^2$ of each other with $\ga^2 \sim
10$: $\ga^{-2} k^2 < |\gth''/\gth| < \ga^2 k^2$. We want to know what the size
of this transition region is in terms of $\eta$, compared to the characteristic
time scale $2\pi/k$ of the solution \eqref{soluearly}. To lowest order in slow
roll $|\gth''/\gth| = a^2 H_k^2 (2 \tge_k + \tget^\parallel_k)$ and $k^2 =
|\gth''_k/\gth_k| = a_k^2 H_k^2 (2 \tge_k + \tget^\parallel_k)$. 
Hence if we define $\get_-$ as the beginning and $\get_+$ as the 
end time of the transition region, we have 
\equ{
\ga^{\pm 1} \equiv
\left | \frac{(\gth''/\gth)_{\get_\pm}}{(\gth''/\gth)_{\get_k}} \right |^{1/2}
= \frac {a(\get_\pm)}{a_k}.
\labl{transregion}
}
By integrating the approximation $\cH(\eta) = a'(\eta)/a(\eta) 
\approx a(\eta) H_k$ around the transition time $\get_k$, we find that 
$1 - a_k/a(\get) = a_k H_k(\get - \get_k)$. 
This leads to the following leading order expression for the ratio of the 
duration of the transition and the period of the oscillation: 
\equ{
\frac {\get_+ - \get_-}{2\gp/k} = 
\frac {k}{a_k H_k} \frac {\ga - \ga\inv }{2\gp} =
\sqrt{2\tge_k + \tget^\parallel_k}\,  \frac {\ga - \ga\inv}{2\gp}.
\labl{transreg}
}
Hence if the slow-roll functions are all individually smaller than $0.01$, 
which is a reasonable assumption as we will show in the examples in section 
\ref{slowroll}, then the transition region does not last longer than 
approximately a tenth of an oscillation period. 

We can also discuss the accuracy of our treatment of the transition between the
small and large wavelength limits in a different way. As we discuss in the next
section, the important quantity is $|D_{\vc{k}}|^2$:
\equ{
|D_{\vc{k}}|^2 
=  \gth_k^2 \lh \lh \frac{\gth'_k}{\gth_k} \rh^2 + k^2 \rh 
=  \gth_k^2 \lh 
\lh \frac{\gth'_k}{\gth_k} \rh^2  + 
\left| \frac{\gth''_k}{\gth_k} \right| \rh  
%=  \frac{\gk^2}{2} \frac{H_k^2}{\tge_k} 
%\lh \frac{k^2}{\cH_k^2} + \lh 1 + \tge_k + \tget^\parallel_k \rh^2 \rh 
= \frac{\gk^2}{2} \frac{H_k^2}{\tge_k},
%\left[
% \bigl( 1 + \tge + \tget^\parallel \bigr)^2 + 
%\bigl|  2\tge + \tget^\parallel + 2 (\tget^\parallel)^2
%- (\tget^\perp)^2 - \tgx^\parallel   \bigr|
%\right] _{t=t_k}.
\labl{DkSR} 
}
where in the last step we have included only the leading order term in slow
roll. We estimate the maximum error in $|D_{\vc{k}}|^2$ as the difference 
between the values of $|D_{\vc{k}}|^2$ determined by matching at $\eta_-$ and
at $\eta_+$:
\equ{
\frac{|D_{\vc{k}}|^2_- - |D_{\vc{k}}|^2_+}{|D_{\vc{k}}|^2_k}
\approx \frac{\lh |D_{\vc{k}}|^2 \rh'_k (\eta_- - \eta_+)}
{|D_{\vc{k}}|^2_k} 
= 2 \lh 2 \tge_k + \tget^\parallel_k \rh \lh \ga - \ga^{-1} \rh
}
to leading order in slow roll, where we made use of \eqref{dereps} and
\eqref{transreg}. From this we conclude that we can indeed only give the
leading  order term for $|D_{\vc{k}}|^2$ in \eqref{DkSR}, since corrections at
the next order are expected for a more accurate treatment of the transition
region. For the single field case an expression for $|D_{\vc{k}}|^2$ that is 
accurate up to and including next-to-leading order terms was obtained in 
\cite{StewartLyth,MartinSchwarz}. 

We finish this section with an argument why the difference between $\eta_k$
and  $\eta_H$ does not matter in the expression for $|D_{\vc{k}}|^2$ to leading
order. Here $\eta_H$ is  the time of horizon crossing, defined by $k^2 =
\cH^2$, which is conventionally used in the literature. On the other hand,
$\eta_k$ is the time when the solution of the differential equation
\eqref{equ2} changes its behaviour, defined by  $k^2 = |\gth''/\gth| = \cH^2
(2\tge+\tget^\parallel+\ldots)$, which we use  to compute $|D_{\vc{k}}|^2$. The
difference between these two expressions for $k^2$ can be quite large during
slow roll. However, just as in the previous paragraph, the relevant quantity is
\equ{
\frac{|D_{\vc{k}}|^2_H - |D_{\vc{k}}|^2_k}{|D_{\vc{k}}|^2_k}
\approx \frac{\lh |D_{\vc{k}}|^2 \rh'_k (\eta_H - \eta_k)}
{|D_{\vc{k}}|^2_k} 
= 2 \, \sqrt{2 \tge_k + \tget^\parallel_k} 
}
to lowest order in slow roll. Here we used that $a_H = a_k \sqrt{2\tge_k +
\tget^\parallel_k}$, as follows from the definitions above, and inserted this
into the expression in the text above \eqref{transreg} to calculate 
$\eta_H-\eta_k$. We see that corrections to  $|D_{\vc{k}}|^2$ because of the
difference between $\eta_k$ and $\eta_H$ are of higher order in slow roll, if
we take the slow-roll functions to be small individually.

\np

\section{Quantum correlation function of the Newtonian potential}
\labl{spectrum}

The quantum correlation function 
\(
\langle \gF (\vc{x}, \get) \gF(\vc{x} + \vc{r}, \get) \rangle 
\)
during inflation with multiple scalar fields is the central object of study in 
this section. In particular, we obtain an expression 
for the correlation function at the end of inflation. Before the actual 
calculation can be performed various questions have to be addressed. 
First of all, can one simply quantize the Newtonian potential? 
Another question is which quantum state is appropriate for the 
computation of the correlator. 

In equation \eqref{newvars} of section \ref{Pertsolv} we have introduced the 
variable $u$  and shown that it decouples from the perpendicular scalar field
fluctations $\Bgd \vc{v}$ if 
corrections of the order of $\sqrt{\tge}\, \tget^\perp$ are 
neglected, and that its equation of motion is then given by \eqref{equ2}. 
However, the Newtonian potential $\gF$ does not correspond to a 
physical degree of freedom within the metric; only the graviton states 
represent physical degrees of freedom and can be quantized in an on-shell 
quantization procedure. (Alternatively, one can use BRS quantization 
to avoid making explicit gauge choices \cite{AndereggMukhanov, HennTeit}.)
The scalar field perturbations $\Bgd\Bgf$ on 
the other hand are physical degrees of freedom, hence they should be 
quantized. But we are not interested in all these scalar perturbations: 
only those that are directly related to the Newtonian potential to this order in
slow roll. The relevant multiple field generalization of the variable $v$
introduced in \cite{Mukhanovetal} is 
\equ{
v \equiv 
\frac {a}{\gk} \lh \vc e_1 \cdot \Bgd\Bgf + \frac {|\Bgf'|}{\cH} \gF \rh 
= 2  \lh u' - \frac {\gth'}{\gth} u \rh, 
\labl{vinu}
}
where we have used \eqref{Einstein0i2} together with the definitions 
\eqref{newvars} and \eqref{equ2}. A slightly different form of this variable
(without the factor $a/\gk$) is sometimes refered to as Mukhanov-Sasaki
variable \cite{Mukhanov,Sasaki}.
The equation of motion for $v$ can be found using the
exact equation of motion for $u$, i.e.\ \eqref{equ} combined with 
\eqref{gthdblgth}, and the expressions for the derivatives of $\cH$ 
\eqref{slowrollrel} and the slow-roll functions \eqref{dereps}:
\equ{
v'' - \gD v - \frac {(1/\gth)''}{1/\gth}v  = 2 \sqrt 2 \, \cH^2 \sqrt{\tge} 
\left[ 
\cH \left( 
\bigl( 3 + \tge + \tget^\parallel \bigr) \tilde \Bget +\tilde \Bgx \right)^\perp
\cdot \Bgd \vc{v} + 
\tilde \Bget \cdot \cD_\get \Bgd \vc{v}
\right]. 
\labl{eqv}
}
The definition of $v$ also includes a term with $\gF$, which ensures that the
equation of motion for $v$ can be written in terms of $v$ and $\Bgd \vc v$
only. (If we were not in the situation where $\Psi=\Phi$, the  definition of
$v$ would contain $\gPs$ instead of $\gF$. This definition is  automatically
gauge invariant, as can be seen from \eqref{deltaphigi}, and  therefore it is
guaranteed that no non-physical degrees of freedom are  quantized.) The scale
factor $a$ is introduced to remove the first derivative term in  the equation
of motion. The result is that at the beginning of inflation, when $k^2 \mg
|(1/\gth)''/(1/\gth)|$ as we shall discuss below, the left-hand side of the 
equation for $v$ is simply the equation of the harmonic oscillator. On the
right-hand side we see that $v$ decouples from $\Bgd\vc{v}$  up to the same
order in slow roll as $u$ does, provided that  $\tilde \Bgx^\perp$ is small as
well. Hence we know how to quantize $v$. Moreover, in this limit of large $k$
the equation of motion for $v$ is equal  to the one for $u$ \eqref{equ2}, up to
terms of order $\sqrt{\tge}\,  \tilde\Bget$ in slow roll. This means that the
quantum operator $\hat u$ can be expanded in terms of the same creation and
annihilation operators as $\hat v$, so that once we have determined the
normalization  by quantizing $v$, we can simply return to the equation for $u$
to determine  the time evolution and compute the correlator at later times.

In the approximation where $v$ decouples from $\Bgd\vc{v}$, 
the action for $v$ given by 
\equ{
S = \int \text{d}^3 \vc{x} \text{d} \get\, {\gk^2} \, 
\left[
\half (v')^2 
+ \half v \Bigl(  \gD + \frac {(1/\gth)''}{1/\gth} \Bigr) v
\right]
\labl{actionu}
} 
gives rise to the equation of motion \eqref{eqv}. 
We continue by canonically quantizing the variable $v$ at the beginning of 
inflation $\get = \get_i$.  
In the quantum theory $\hat v$ and its canonical momentum 
$\hat \gp = \gk^2 \hat v'$ satisfy the commutation relation 
\equ{
[ \hat v(\vc{x}, \get),\hat \gp(\vc{y}, \get)] = i \gd(\vc{x} - \vc{y}).
}
At the beginning of inflation the field $\hat v$ can be expanded as follows:
\equ{
\hat v(\vc{x}, \get) = 
\int \frac{\text{d}^3 \vc{k}}{(2\gp)^{3/2}\, \sqrt{2k}\, \gk}
\left\{
v_{\vc{k}}^{\;}(\get) e^{-i \vc{k} \vc{x}} \,\hat b_{\vc{k}}^\dag + 
v^*_{\vc{k}}(\get) e^{i \vc{k} \vc{x}} \, \hat b_{\vc{k}}^{\;}
\right\},
}
where the time independent creation and annihilation operators 
$\hat b^\dag_{\vc{k}}$ and $\hat b^{\;}_{\vc{k}}$ obey the 
commutation relation 
\equ{
[ \hat b_{\vc{k}}^{\;}, \hat b^\dag_{\vc{l}}] = \gd(\vc{k} - \vc{l}).
}
In order for this expansion to be valid
the dimensionless mode functions $v^{\;}_{\vc{k}}(\get)$ and  
$v^{*}_{\vc{k}}(\get)$ have to satisfy the Wronskian condition
\equ{
W(v^*_{\vc{k}}(\get), v_{\vc{k}}(\get)) = 
v^*_{\vc{k}}(\get) v^\prime_{\vc{k}}(\get) 
- v^{*\prime}_{\vc{k}}(\get) v_{\vc{k}}^{\;}(\get) = i\, 2k,
\labl{Wronskianv}
}
where the $v_{\vc{k}}$ only depend on the length $k$ of $\vc{k}$.
It can be checked that the equation of motion implies 
that $\smash{\dd[]{\get}} W(v^*_{\vc{k}}, v_{\vc{k}}) = 0$, 
so that the quantization procedure is consistent with the 
equation of motion; this is guaranteed by using canonical quantization. 
At the initial stages of inflation the scales $k^2$ that 
are observable in the CMBR today are much, much larger than the 
horizon: $k^2 \mg \cH^2$. We assume that this implies 
that $k^2$ is also very large  
compared to the other relevant quantities $|\gth''/\gth|$, $(\gth'/\gth)^2$ 
and $|(1/\gth)''/(1/\gth)|$, since they are all proportional to $\cH^2$, up to 
factors of order unity or slow roll. Therefore, when inflation starts 
the variable $v$ can be expanded in terms of the independent 
solutions $e_{\vc{k}}(\get)$ and $e^*_{\vc{k}}(\eta)$ defined in 
\eqref{soluearly}. As explained below \eqref{eqv}, we can expand $\hat u$ in 
the same way. Introducing the notation 
\equ{
\gTh \equiv \pmtrx{1&0\\0&-1},
\qquad
\hat B_{\vc k} = \pmtrx{\hat b^\dag_{\vc k} \\ \hat b_{\vc k}},
\qquad
E_{\vc k} = \pmtrx{ e_{\vc k} & e^*_{\vc k} }
\qquad\Rightarrow\qquad
E_{\vc k}' = i k E_{\vc k} \gTh, 
\labl{defEB}
}
we may write $\hat v$ and $\hat u$ as
\equ{
\hat v(\vc{x}, \get) = 
\int \frac{\text{d}^3 \vc{k}}{(2\gp)^{3/2}\, \sqrt{2k}\, \gk}
E_{\vc k} R_{\vc k} e^{- i \vc k \vc x\, \gTh} \hat B_{\vc k},
\qquad
\hat u(\vc{x}, \get) = 
\int \frac{\text{d}^3 \vc{k}}{(2\gp)^{3/2}}
E_{\vc k} U_{\vc k} e^{- i \vc k \vc x\, \gTh} \hat B_{\vc k}.
}
Combining the Wronskian condition for $e_{\vc{k}}$ \eqref{Wronskianupm}, 
written as $W(E_{\vc k}) = E_{\vc k} \gTh E^\dag_{\vc k} = i \, 2k$, 
with the Wronskian condition for $v_{\vc{k}}$ \eqref{Wronskianv}, 
$W(E_{\vc k} R_{\vc{k}}) = E_{\vc k} R_{\vc{k}} \gTh R^\dag_{\vc{k}} 
E^\dag_{\vc k} = i \, 2k$, we see that
$R_{\vc k}$ is an element of the non-compact unitary group 
$U(1,1)$ defined by 
\equ{
R^{\;}_{\vc k} \gTh R^\dag_{\vc k} = \gTh.
}
In addition the Hermiticity of $\hat{v}$ implies that $R_{\vc{k}}$ is an 
element of $SU(1,1)$.
The matrix $R_{\vc k}$ represents a 
Bogolubov transformation \cite{BirrellDavies}.
Using the relation \eqref{vinu} between $v$ and $u$, \eqref{defEB} for 
$E'_{\vc{k}}$, and the fact that $k \mg |\gth'/\gth|$, the expansion 
in quantum mechanical operators leads to
\equ{
U_{\vc k} = \frac{-i}{(2k)^{3/2} \gk} \, \gTh R_{\vc k}. 
}

We now derive a general compact expression for the expectation 
value of the correlator 
\(
\langle \hat u (\vc{x}, \get) \hat u^\dag (\vc{x} + \vc{r}, \get) \rangle_\gr
\)
computed in an arbitary state that is represented by the density matrix 
$\hat \gr$. 
We assume that expectation values  
$\langle \hat b_{\vc k} \hat b_{\vc l}  \rangle_\gr = 0$, etc., 
if $\vc k \neq \vc l$. 
Then
\equ{
\langle \hat B_{\vc k} \hat B^\dag_{\vc l} \rangle_\gr = 
\pmtrx{ \langle N_{\vc k} \rangle_\gr & 
\langle \hat b_{\vc k}^2 \rangle_\gr^* 
\\ 
\langle \hat b_{\vc k}^2 \rangle_\gr
& 1 + \langle \hat N_{\vc k} \rangle_\gr }
\gd(\vc k - \vc l),
}
with the number operator 
$\hat N_{\vc k} =  \hat b^\dag_{\vc k} \hat b^{\;}_{\vc k}$. 
The two-point correlator becomes
\equ{
\langle \hat u (\vc{x}) \hat u^\dag (\vc{x} + \vc{r}) \rangle_{\hat \gr} = 
\int \frac{\text{d}^3 \vc{k}}{(2\gp)^{3}} \, 
E_{\vc k} \, U_{\vc k} \, e^{- i \vc k \vc x\, \gs} 
\pmtrx{ \langle \hat N_{\vc k}\rangle_\gr & 
\langle \hat b_{\vc k}^2 \rangle_\gr^*  \\ 
\langle \hat b_{\vc k}^2 \rangle_\gr & 
1 + \langle \hat N_{\vc k} \rangle_\gr }
e^{i \vc k (\vc x + \vc r)\, \gs} \, 
U^\dag_{\vc k} \, E^\dag_{\vc k}.
}

We illustrate this expression with two examples. 
The simplest state to consider is a conformal vacuum state density matrix 
$\hat \gr_0 = | 0 \rangle \langle 0 |$ where for all $\vc k$ the state $| 0 \rangle$
is annihilated by $\hat b^{\;}_{\vc k}$. The correlator then takes the form 
\equ{
\langle \hat u (\vc{x}) \hat u^\dag (\vc{x} + \vc{r}) \rangle_{0}
= \int \frac{\text{d}^3 \vc{k}}{(4\gp\, k)^{3} \gk^2}  
\left| 
(R_{\vc k})_{22} \, e^*_{\vc k}
- (R_{\vc k})_{12}  \, e_{\vc k}
\right|^2 \,
e^{-i \vc k \vc r}.
\labl{corruvac}
}
This shows that $R_{\vc k}$ represents a Bogolubov 
transformation. In other words, $R_{\vc k}$ measures the alignment 
of the expansion of the field $\hat u$ in terms of creation and annihilation 
operators ($\hat b^\dag_{\vc k}$ and $ \hat b_{\vc k}$) with respect to the 
Lorentz conformal vacuum. 
In the Lorentz conformal vacuum all matrices $R_{\vc k}$ are equivalent to 
the identity, so that the field expansion takes the familiar Lorentz-invariant 
form 
\equ{
\hat u(\vc{x}, \get) = 
\int \frac{\text{d}^3 \vc{k}}{(4\gp\, k)^{3/2} \gk}
\left\{
e^{i(k \get - \vc{k} \vc{x})} \, \hat b_{\vc{k}}^\dag + 
e^{-i(k\get - \vc{k} \vc{x})} \, \hat b_{\vc{k}}^{\;}
\right\}.
}
Here we used the definition of the functions $e_{\vc k}$ given in  
\eqref{soluearly} for modes with $k^2 \mg |\gth''/\gth|$.  
The canonical Hamiltonian
$\hat{H}$ associated with the action \eqref{actionu} is given by
\equ{
\hat{H} = \int \d^3 \vc{k} \, \frac{k}{2} \left [ \sinh(2\vartheta)
\lh e^{i 2\gd} \, \hat b_{\vc{k}}^\dag \hat b_{-\vc{k}}^\dag 
+ \mathrm{h.c.} \rh
+ \cosh(2\vartheta) \, (2 \hat N_{\vc{k}} + 1) \right ],
}
where we wrote
\equ{
R_{\vc{k}} = \pmtrx{\cosh \vartheta \: e^{i(\varphi+\gd)} &
\sinh \vartheta \: e^{i(\varphi-\gd)} \\
\sinh \vartheta \: e^{-i(\varphi-\gd)} &
\cosh \vartheta \: e^{-i(\varphi+\gd)} },
}
which is a representation of the most general form for an element of $SU(1,1)$.
Here $\vartheta$, $\varphi$ and $\gd$ are real, and in general depend on
$\vc{k}$. We see that for $\vartheta=0$ there is no pair creation of
particles and anti-particles (the $\hat b_{\vc{k}}^\dag \hat b_{-\vc{k}}^\dag$
term drops out), and $\cosh(2\vartheta)$ takes its minimum value at 
$\vartheta=0$. Hence choosing the Lorentz alignment $\vartheta=0$ seems to lead
to a configuration with the usual vacuum properties.

Our second example is a thermal state of temperature $1/\gb$ in Planckian units 
characterized by 
\equ{
\gr_\gb \propto e^{- \gb \hat H}, 
\qquad\qquad
\hat H = \int \d^3 {\vc k} \, k \, \hat N_{\vc k},
\qquad\qquad
\langle \hat N_{\vc{k}} \rangle_{\gr_\gb} 
= \frac 1{e^{\gb k} - 1}.
\labl{finitetemp}
}
Here we chose the Lorentz alignment and neglected the (infinite) zero-point 
energy, which is irrelevant for the definition of the density matrix.
In the thermal state we have that 
$\langle \hat b_{\vc k}^2 \rangle_{\gr_\gb} = 0$, hence we find 
for the thermal correlator 
\equ{
\langle \hat u(\vc{x}, \get) \hat u(\vc{x} + \vc{r}, \get) \rangle_{\gr_\gb} = 
\int \frac{ \text{d}^3 \vc{k}} {(4\gp\, k)^3 \gk^2}  
| e_{\vc{k}}(\get) |^2 \, 
\left(
1 + 2 \langle \hat N_{\vc k} \rangle_{\gr_\gb}
\right) \, e^{-i \vc{k} \vc{r}},
}
where the occupation number $\langle \hat N_{\vc k} \rangle_{\gr_\gb}$ is given 
in \eqref{finitetemp}.

Next we argue why taking the vacuum state $| 0 \rangle$ at the beginning  of
inflation  is a reasonable assumption for the calculation of the density
perturbations that we can observe in the CMBR today. Even though perturbations
in the  CMBR have long wavelengths now, they had very short wavelengths before
they went through the horizon during inflation. Therefore,  their scale $k$ at
the beginning of inflation at $t_i$ is much larger  than the Planck scale. It
seems a reasonable assumption that modes with momenta very much larger than 
the Planck scale are not excited at $t_i$, so that for these modes the vacuum
state is a good assumption.\footnote{There could be a problem with this
approach,  because our knowledge of physics beyond the Planck scale is
extremely poor.  In particular, the dispersion relation $\go(\vc{k})=k$ that we
used implicitly might not be valid for large $k$: there might be a cut-off for
large momenta.  For a discussion of this trans-Planckian problem and possible
cosmological consequences see \cite{MartinBrandenberger}.}

This argument becomes more convincing if we put in some numbers. Assume that
inflation starts around the  Planck time, $t_i= t_P$,  when the horizon is
naturally of the order of the  Planck scale: $H \approx \gk\inv$.  Of course we
do not know the exact quantum state at the Planck time, but as a first
indication we take a thermal state with a temperature of the order of  the
Planck energy: $\gb \approx 1$.\footnote{Other non-vacuum initial states and
possible observable effects have been investigated in \cite{Martinetal}.}  The
conformal Hubble parameter is given by $\cH = \gk a_i e^N H \approx a_i e^N$
with $N=\int_{t_i}^t H \d t$, where  in the last step we made the approximation
of a constant $H$ during (the beginning of) inflation.  Now we evaluate $\cH$
at the time when the scale $k$ goes through the horizon, $k=\cH$. Suppose that
inflation started just before this happens, say $N \approx 2$, and that $a_i
\approx 1$ (the universe should at least be a Planck length at the Planck time,
and a larger value only makes  the argument stronger). Even then we find that
the thermal  correction is already very small:  $2 \langle \hat N_{\vc k}
\rangle_{\gr_\gb}\approx 10^{-3}$.  In physically more realistic situations the
number of e-folds $N$ can easily  be of the order of $100$, so that this
thermal effect is completely negligible  because $\exp (\exp 100))$ gives a
huge suppression. 

As we are making a claim concerning physics at a time when we 
do not have any solid theoretical or experimental data, we have 
to make sure that our statement does not depend strongly on the 
precise choice of the initial time $t_i$ of inflation. However, because 
$|\gth''/\gth| \ll k^2$ the field equation \eqref{equ2} of $u$ is 
a plain wave equation, it follows that the Bogolubov transformation 
between the Fock spaces at different initial times can be neglected. 
This implies that in our statement there is no fine-tuning problem 
associated with the starting time of inflation. 

In the remainder of this section we compute the vacuum correlator for the
gravitational potential $\Phi$, derive expressions for the amplitude and the
slope of the density perturbation spectrum, and compare our results with the 
literature.
Using the Lorentz alignment for the vacuum, the vacuum correlation function 
for $\Phi=\gk^3 |\dot{\Bgf}| u$ can be calculated from \eqref{corruvac} and
\eqref{longu}:
\equ{
\langle \Phi(\vc x, t) \Phi(\vc x + \vc r, t) \rangle_0 = 
\half 
\int \frac{ \text{d}^3 \vc{k}} {(2\gp\, k)^3}  
\, {|D_{\vc k}|^2}\, 
\lh 1 - \frac{H}{a} \int_{t_k}^t a(\gt) \, \d \gt \rh^2
 \,e^{-i \vc{k} \vc{r}},
}
where we have used that for large time $t \mg t_k$ the first term in 
\eqref{longu} can be neglected because of the $1/a$ suppression. 
The only model dependence resides in the norm of the coefficient 
$|D_{\vc{k}}|^2$, given in \eqref{DkSR}. 

Using the fact that $|D_{\vc{k}}|^2$ only depends on the length of $\vc{k}$,
we can perform the integration over the angles and obtain:
\equ{
\langle \Phi(\vc x, t) \Phi(\vc x + \vc r, t) \rangle_0 = 
\int_0^\infty \frac{\d k}{k} \frac{\sin kr}{kr} \, |\gd_{\vc{k}}(t)|^2,
}
where
\equ{
|\gd_{\vc{k}}(t)|^2 = \frac{1}{4\pi^2} |D_{\vc{k}}|^2 
\lh 1 - \frac{H}{a} \int_{t_k}^t a(\gt) \, \d \gt \rh^2.
}
To obtain an estimate of the size of this quantity, we first observe that 
\equ{
1 - \frac{H}{a} \int_{t_k}^t a(\gt) \, \d \gt =
\lh \frac{1}{a} \int_{t_k}^t a(\gt) \, \d \gt \rh^{\boldsymbol{\cdot}}. 
}
Because $H(t)$ and $a(t)$ are positive functions, we see immediately that 
the left-hand side has to be smaller than
or equal to one for $t \geq t_k$.
On the other hand, as long as $a(t)$ is positive, and does not grow faster than
exponentially, the term within brackets on the right-hand side will be a
non-decreasing function of $t$, so that its time derivative is non-negative.
Hence for all cases of interest
\equ{
0 \leq \lh \frac{1}{a} \int_{t_k}^t a \, \d \gt \rh^{\boldsymbol{\cdot}} \leq 1
\qquad\Rightarrow\qquad
|\gd_{\vc{k}}(t)|^2 \leq \frac{1}{4\pi^2} |D_{\vc{k}}|^2.
}

As mentioned in the introduction, we extrapolate our result right to the time
of recombination for the purpose of comparing it with the results in the 
literature. In \cite{Wandsetal} it is shown that this is justified for
adiabatic perturbations, which are the only ones we consider here. However, a
full discussion of the validity of this extrapolation is beyond the scope of
this paper. At the time of recombination during matter domination $a(t) \propto
t^{2/3}$,  which leads to
\equ{
|\gd_{\vc{k}}|^2 = \frac{9}{25} \frac{1}{4\pi^2} |D_{\vc{k}}|^2.
}
To leading order in slow roll this is exactly the same result as that obtained
in \cite{LiddleLyth}, if one takes into account that the $\gd_H^2$ defined in
that paper equals $\frac{4}{9} |\gd_{\vc{k}}|^2$. The calculation of the slope
of the spectrum is analogous to the one presented in \cite{LiddleLyth} and
gives to leading order in slow roll:
\equ{
n-1 \equiv \dd[\ln |\gd_{\vc{k}}|^2]{\ln k} = -4 \tge_k - 2 \tget^\parallel_k
= -6 \ge_k + 2 \eta_k,
}
where in the last step we took the single field limit and switched to the
conventional slow-roll parameters \eqref{convsrpar} for the purpose of
comparison. We see that in that case the result is identical to the one in
\cite{LiddleLyth}. Comparing with the result in \cite{SasakiStewart} we see,
after rewriting it in terms of our slow-roll functions, that all corrections
are indeed of higher order.

\np

\section{Slow roll with multiple scalar fields}
\labl{slowroll}

\subsection{Slow roll on a flat manifold}
\labl{SlowRollFlat}

In this section we consider some examples of slow-roll inflation with scalar 
fields living on a flat manifold. Since all flat 
manifolds are locally isomorphic to a subset of $\Real^N$, 
we assume that the $N$ scalar fields live in the $\Real^N$ 
themselves. In particular, we use the standard basis for 
$\Real^N$. 
The (zeroth order) slow-roll equation of motion and Friedmann equation 
for the background quantities are given by
\equ{
\Dot \Bgf = - \frac 2{\sqrt{3}\, \gk} \, \Bder^T \sqrt{V(\Bgf)},
\qquad
H = \frac{\gk}{\sqrt{3}} \, \sqrt{V(\Bgf)}.
\labl{emSRflat}
}
We make use of the hat to indicate a unit vector: 
$\hat{\Bgf}\equiv \Bgf/\gf$, with $\gf \equiv \sqrt{\Bgf^T \Bgf}$ the length of
the vector $\Bgf$.
In the first subsection we consider a quadratic potential where all 
scalar field components have equal masses, while in the second 
subsection we focus on  the more complicated case of a quadratic 
potential with an arbitrary mass matrix.

\subsubsection{Scalar fields with identical masses on a flat manifold}
\labl{eqmassflat}

In this example all masses are assumed to be equal to $\gk\inv m$, so that 
the mass matrix is proportional to the identity matrix and the 
potential reads 
\(
V = \smash{\half} \gk^{-2} m^2 \gf^2.
\)
The mass parameter $m$ is given in units of the reduced Planck mass 
$\gk\inv$. 
The slow-roll equation of motion for the background fields 
simplifies to
\equ{
\dot{\Bgf} = \dot{\gf} \, \hat{\Bgf} + \gf \, \Dot{\hat{\Bgf}} 
= - \sqrt{\frac{2}{3}} \, \frac{m}{\gk^2} \, \hat{\Bgf} 
\qquad \Ra \qquad
\dot{\gf} = - \sqrt{\frac{2}{3}} \, \frac{m}{\gk^2}
\qand 
\Dot{\hat{\Bgf}}=0.
}
Here we have used the fact that $\hat \Bgf$ and $\Dot{\hat{\Bgf}}$ are 
perpendicular,  as can  be seen by 
differentiating the relation $\hat{\Bgf}^T \hat{\Bgf} = 1$.
This means that the direction of $\Bgf$ is fixed in time; 
only its magnitude changes. The scalar equation can of course
be solved easily, and we obtain 
\equ{
\Bgf(t) = \lh 1 - \frac{t}{t_\infty} \rh \Bgf_0,
\qquad\mbox{with}\qquad
t_\infty = \sqrt{\frac{3}{2}} \frac{\gk^2 \gf_0}{m}, 
}
where we used the initial condition $\Bgf(0) = \Bgf_0$. 
Here $t_\infty$ is the time when $\Bgf=0$ if slow roll would be valid 
until the end of inflation. 

Using this solution, we calculate the Hubble parameter $H$ and the number of
e-folds $N$:
\equ{
H(t) = \frac{m}{\sqrt{6}} \, \gf = 
\frac{m \gf_0}{\sqrt{6}} \lh 1 - \frac{t}{t_\infty} \rh,
\qquad
N(t) = \int_0^t H \d t' = 
N_\infty - \frac{1}{4} \gk^2 \gf_0^2 \lh 1 - \frac{t}{t_\infty} \rh^2,
}
where $N_\infty = \frac{1}{4} \gk^2 \gf_0^2$.
Next we calculate the slow-roll functions. Since $\Bgf(t)$ is linear in time,
$\tget^\parallel$ and $\tget^\perp$  are zero to this order in slow roll. This
implies that to this order the decoupling  of the Newtonian potential from the
scalar field perturbations is exact, see  \eqref{matrixeq}.  For $\tge$ we
find 
\equ{
\tge = (H\inv)^{\boldsymbol{\cdot}} = 
\frac{2}{\gk^2 \gf_0^2} \lh 1 - \frac{t}{t_\infty} \rh^{-2}.
}
Clearly, $\tge$ becomes infinite when $t \ra t_\infty$, which is in 
contradiction with the bound $\tge < 3$ derived in subsection
\ref{slowrollsec}. But of course slow roll has certainly stopped 
when  $t \geq t_1 \equiv t_\infty -\sqrt{3}\, \gk/m$, because then 
$\tge \geq 1$, so that results obtained from equations valid only within slow 
roll cannot be trusted. Notice that we do not have a
slow-roll period at all  if $\gf_0 \leq \sqrt{2}/\gk$, since then
$t_1 \leq 0$, so that $\tge$ is never smaller than $1$.
We regard the quantity $N_\infty$ as the number of
e-folds at the end of inflation. This might seem questionable, as $t_1$ is a
better candidate for the end time of (slow-roll) inflation than $t_\infty$. 
However, the difference in the number of e-folds using $t_\infty$ and $t_1$ is 
small: $N_\infty - N(t_1) = \half$, so that we can safely use 
$N_\infty$ as a good approximation for the number of e-folds at the end of 
inflation. 

We finish by calculating the model dependent factor $|D_{\vc{k}}|^2$, which we 
need to compute the correlation function of $\Phi$ (see section
\ref{spectrum}). To this end we must determine the time $t_k$ when a mode
function $u_{\vc{k}}$ changes  its behaviour (see section \ref{Pertsolv}). We
are especially interested in those scales that are observable in the CMBR. As
mentioned in the introducion of this paper, the corresponding mode functions
change behaviour in a small interval about 60 e-folds before the end of
inflation. Hence we consider $N_k \approx 60$ to be a fixed quantity, and 
determine $t_k$ by means of the definition $N_k = N_\infty - N(t_k)$.  
We find
\equ{
\frac{t_k}{t_\infty} = 1 - \sqrt{\frac{N_k}{N_\infty}}
\qquad\Rightarrow\qquad
\tge_k = \frac{1}{2 N_k},
}
so that slow roll is still a good approximation at time $t_k$.
To leading order in slow roll we obtain the following expression for
$|D_{\vc{k}}|^2$: 
\equ{
|D_{\vc{k}}|^2 =  \frac{\gk^2}{2} \frac{H_k^2}{\tge_k}
= \frac{2}{3} \, m^2 \, N_k^2.
\labl{Dkflatcentpot}
}

\subsubsection{Scalar fields with a quadratic potential on a flat manifold}
\label{flatmasses}

Now we consider a more general symmetric mass matrix $\gk\inv \Bsfm$ 
in the potential. 
It does not necessarily have to be diagonalized, but because it is 
symmetric we can always bring it in diagonal form.
As a further assumption we take all eigenvalues to be positive, 
otherwise the potential would not be bounded from below.
The potential is denoted by $V_2$ and given by
\equ{
V_2 = \half \gk^{-2} \, \Bgf^T \Bsfm^2 \Bgf,
\labl{sqpot}
}
so that the slow-roll equation of motion (\ref{emSRflat}) reduces to
\equ{
\Dot \Bgf  = -\sqrt{\frac {2}{3}}  \, \frac{1}{\gk^2} \, 
\frac {1} {\sqrt{ \Bgf^T \Bsfm^2 \Bgf }} \, \Bsfm^2 \Bgf.
\labl{dvBgf}
}
The solution of this vector equation can be written in terms of one 
dimensionless scalar function $\gps(t)$ as
\equ{
\Bgf(t) = e^{ - \half \Bsfm^2 \gps(t)} \Bgf_0.
\labl{ansatz_Bgf}
}
Here $\Bgf_0 = \Bgf(0)$ is the initial starting point of the field $\Bgf$, 
which implies that $\gps(0) = 0$. In other words, we have determined 
the trajectory that the field $\Bgf$ follows through field space starting 
from point $\Bgf_0$. 

An important role in our further analyses is played by the functions 
$F_n$, defined by
\equ{
F_n 
%= \frac {\Bgf^T \Bsfm^{2n} \Bgf}{\gf_0^2}
= \frac {\Bgf_0^T \Bsfm^{2n} e^{-\Bsfm^2 \gps}\Bgf_0}{\gf_0^2},
\labl{defF}
}
with $\gf_0$ the length of $\Bgf_0$: $\gf_0^2 = \Bgf_0^T \Bgf_0$.
%The first definition does not depend on any explicit 
%representation of $\Bgf$ as a function of time, 
%hence it can also be used in more complicated situations.
%In the second expression we have inserted the ansatz
%(\ref{ansatz_Bgf}) for the trajectory.
The functions $F_n(\gps)$ are positive and monotonously decreasing
for all $\gps$, tending to zero in the limit $\gps\rightarrow\infty$, 
because we have assumed that all mass eigenvalues are positive. 
Using these definitions we see that the function $\gps(t)$ is determined by 
the following equations:
\equ{
\dot{\gps} = \sqrt{\frac {2}{3}} \frac{2}{\gk^2 \gf_0}
\frac{1}{\sqrt{F_1(\gps)}},
\qquad\qquad
\gps(0)=0.
\labl{eqgps}
} 
Notice that $\gps$ is always non-negative.
The functions $F_n$ do not depend on the length of $\Bgf_0$, only on its 
direction, as can be seen from the definition (\ref{defF}). 
This implies that the only dependence on the 
length $\gf_0$ in \eqref{eqgps} is in the factor $2/(\gk^2 \gf_0)$, so that 
it can be absorbed by a redefinition of the 
time variable only.

Next we discuss some additional properties of the functions $F_n$. 
The definition of $F_n$ can also be written as
\equ{
F_n = \frac {\Bgf_0^T e^{-\half \Bsfm^2 \gps} \Bsfm^{n-p} \Bsfm^{n+p} 
e^{-\half \Bsfm^2 \gps} \Bgf_0}{\gf_0^2},
}
for any integer $-n \leq p \leq n$. Using the Green-Schwarz inequality 
$ ( \vc A^T \vc B)^2 \leq (\vc A^T \vc A) (\vc B^T \vc B)$  
for arbitrary vectors $\vc A$ and $\vc B$, we obtain 
\equ{
F_n^{\; 2} \leq  F_{n-p} \; F_{n+p}.
\labl{GrSchwF}
}
From the definition of the $F_n$ we also see that 
\equ{
\dd{\gps} F_n (\gps) = - F_{n+1}(\gps).
\labl{derF}
}

We can express many important quantities in the functions $F_n$.
The Friedmann equation (\ref{emSRflat}) for $H$ simplifies to
\equ{
H = \frac{\gf_0}{\sqrt{6}} \, \sqrt{F_1(\gps)}.
\labl{Hgps}
}
Formally integrating equation (\ref{eqgps}) for $\gps$, 
we find how long it takes to go from $\gps=0$ to $\gps$ in slow roll:
\equ{
t(\gps) = \sqrt{\frac 38} \gk^2 \gf_0  
\int_0^{\gps} \text{d} \gps' \sqrt{F_1(\gps')} .
\labl{timegps}
} 
The number of e-folds $N = \int H \d t$ can
be interpreted as a function of $\gps$ given by
\equ{
N(\gps) = \frac{\gk^2 \gf^2_0}{4} 
\int_0^{\gps} \text{d} \gps' F_1(\gps') = 
N_\infty \left( 1 - F_0(\gps) \right),
\labl{efolds_flat}
}
where we combined (\ref{Hgps}) and (\ref{timegps}), and used (\ref{derF}) 
to perform the integration. The number of e-folds in the limit $\gps\ra\infty$
is given by $N_\infty = \frac{1}{4} \gk^2 \gf_0^2$, which can be interpreted as
an upper limit for the total number of e-folds during slow-roll inflation.
Finally, the slow-roll functions \eqref{slowrollfun2} can now be written as
\equ{
\tge = \frac {2}{\gk^2 \gf_0^2} \frac{F_2}{F_1^2},
\qquad
\tget^\parallel = - \frac {2}{\gk^2 \gf_0^2} 
\frac {F_3 F_1 - F_2^2}{F_1^2 F_2}, 
\qquad
\tget^\perp = 
\frac {2}{\gk^2 \gf_0^2} 
\frac{\sqrt{F_4 F_2 - F_3^2}} {F_1 F_2}.
\labl{slowrollfun_flat}
}
Using the Green-Schwarz inequality (\ref{GrSchwF}) we see that
$\tget^\parallel$ is always negative, while $\tget^\perp$ is real, as it should
be. Observe that if $\Bsfm$ is proportional to the identity, the inequality is
saturated and $\tget^\parallel$ and $\tget^\perp$ are zero. This is in
agreement with the results of  the previous subsection.  Since the functions
$F_n(\gps)$ are independent of $\gf_0$, this dependence enters only in the
prefactors of the expressions for $t(\gps)$, $N(\gps)$ and the slow-roll 
functions.

Before going on to discuss estimates for the functions $F_n$, we need to
introduce some additional notation. We define a semi-positive definite matrix 
norm:
\equ{
%\langle \mx A, \mx B \rangle = 
%\frac{ \Bgf_0^T \, \mx A^T \mx B \;\Bgf_0}{\Bgf_0^T \Bgf_0},
%\qquad
|| \mx A ||^2 
% = \langle \mx A, \mx A \rangle 
= \frac {| \mx A \Bgf_0 |^2}{\gf_0^2},
\labl{flat_matrix_inp}
}
for any arbitrary $N\times N$-matrix $\mx A$. 
The reason that $||\cdot||$ does not define a regular norm 
is that $|| \mx A ||^2 = 0$ does not imply that 
$\mx A = 0$; we can only infer that $ \mx A \Bgf_0 = 0$.
Indeed, if $\mx A$ has determinant zero and $\Bgf_0$ is 
one of $\mx A$'s zero modes, $\mx A \Bgf_0 = 0$ is satisfied without 
$\mx A$ being the zero matrix. 
With this norm the definition of $F_n(\gps)$ can also be written as
\equ{
F_n(\gps)= || \Bsfm^n e^{- \half \Bsfm^2\gps}||^2.
}
We order the eigenvalues of $\Bsfm^2$ from smallest to largest,
$m_1^2 < m_2^2 < \ldots < m_\ell^2$. Here we look only at distinct eigenvalues,
so that $\ell$ is smaller than $N$ if there are degenerate eigenvalues.
The projection operator $\mx{E}_n$ projects on the eigenspace 
with eigenvalue $m_n^2$. These operators are mutually orthogonal and sum 
to the identity: $\sum \mx E_n = \Id$. 
The norm of these projection operators satisfies $||\mx{E}_n||\leq 1$.

Above we have been able to write all kinds of important  quantities for the
slow-roll period in terms of the functions  $F_n(\gps)$. But these functions
are rather complicated as they depend both on an (exponentiated) mass matrix
$\Bsfm^2$ and  on the direction of the initial vector $\Bgf_0$.  Since in the
CMBR we cannot see further back than about the last 60 e-folds of inflation,
while the total amount of inflation is generally much larger, we are often only
interested in the asymptotic behaviour  of the functions $F_n(\gps)$ for large
$\gps$.  Below we will show that the asymptotic behaviour is indeed a good
approximation for the time interval during inflation that is (indirectly)
observable through the CMBR,  but first we concentrate on the asymptotic
expressions themselves.

As can be seen from the definition of $F_n$ in (\ref{defF}), in the limit
$\gps\rightarrow\infty$ the smallest mass eigenvalue will start to dominate. We
denote the smallest eigenvalue by $\gm$, $\gm\equiv m_1$, while the ratio of
the next-to-smallest and smallest masses squared is called $\gr$:  $\gr \equiv
m_2^2/m_1^2 >1$. Furthermore, the operator $\mx{E}\equiv\mx{E}_1$ projects on
the eigenspace of the smallest  eigenvalue,\footnote{Here we assumed that
$||\mx{E}_1||\neq 0$. If $||\mx{E}_1||$  is zero, it means that $\Bgf_0$ has no
component in the subspace corresponding with this eigenvalue.  As can be seen
from the differential equation for this  quadratic case (\ref{dvBgf}), this
means  that $\Bgf$ will never obtain a component in the directions 
corresponding to this subspace. Hence we remove these directions from the
problem, consider $\Bgf$ to be a vector of appropriate (lower) dimension, and 
take $m_1$ to be the smallest remaining eigenvalue, etc.} 
and we define $\gch \equiv ||\mx{E}_2||^2/||\mx{E}_1||^2$. 
Using these definitions, we find the following asymptotic behaviour 
for the functions $F_n(\gps)$ in the limit $\gps\rightarrow\infty$:
\equ{
F_n \ra ||\mx{E}||^2 \gm^{2n} e^{- \gm^2 \gps} 
\left(
1 + \gch \gr^n e^{-(\gr-1)\gm^2\gps}
\right) 
\ra ||\mx{E}||^2 \gm^{2n} e^{- \gm^2 \gps},
\labl{estF}
}
where the first limit contains both leading and next-to-leading order terms,
while the second contains only the leading order term.
Both these asymptotic expressions for $F_n$ are needed to obtain the
non-vanishing leading order behaviour of ratios and differences of 
ratios of the functions $F_n$:
\equ{
\frac {F_p}{F_q} \ra \gm^{2(p-q)}, 
\qquad
\frac {F_{n+1}}{F_n} - \frac {F_n}{F_{n-1}} \ra
\gm^2 \gr^{n-1}(\gr -1)^2 \gch e^{-(\gr-1) \gm^2 \gps}.
\labl{estFF}
}
Using these expressions we find the asymptotic behaviour for the Hubble
parameter (\ref{Hgps}) and the number of e-folds (\ref{efolds_flat}),
\equ{
H(\gps) \ra 
\frac{\gf_0 \, \gm ||\mx{E}||}{\sqrt{6}} \, e^{-\half\gm^2 \gps},
\qquad\qquad
%t(\gps) \ra \sqrt{\frac{3}{2}} \, \gk^2 \gf_0 \, \frac{||\mx{E}||}{\gm}
%\lh 1 - e^{-\half \gm^2 \gps} \rh, 
%\non 
%\\
N(\gps) \ra 
N_\infty \lh 1-||\mx{E}||^2 e^{-\gm^2 \gps} \rh. 
\labl{Nlim}
}
The asymptotic behaviour of the slow-roll functions 
(\ref{slowrollfun_flat}) is given by
\equa{
\tge & \ra \frac {2}{\gk^2 \gf_0^2} \frac 1{||\mx{E}||^2} \, e^{\gm^2 \gps}, &&
\tget^\parallel \ra - \frac {2}{\gk^2 \gf_0^2} \frac {\gch}{||\mx{E}||^2} 
\, \gr (\gr -1)^2 e^{-(\gr-2) \gm^2 \gps}, \non\\
\tget^\perp & \ra \frac {2}{\gk^2 \gf_0^2}  \frac {\sqrt {\gch}}{||\mx{E}||^2} 
\, \gr (\gr -1) e^{-\half (\gr-3)\gm^2 \gps}.&&
}
Notice that $\tget^\parallel$ goes to zero for $\gr>2$, while for $\gr<2$ it
diverges.  The same holds true for $\tget^\perp$, but there the critical value
is $\gr=3$. Since $\gr>1$ by definition, the slow-roll function $\tge$ always 
grows faster than $\tget^\parallel$ and $\tget^\perp$   in the limit
$\gps\ra\infty$. As we discussed in section \ref{slowrollsec}, however, it is 
really the combinations $\sqrt{\tge}\, \tget^\parallel$ and $\sqrt{\tge}\, 
\tget^\perp$ that determine whether slow roll is valid or not. 
Their critial values are $\gr = 5/2$ and $\gr = 4$, respectively. 

We finish this subsection by calculating the model dependent factor 
$|D_{\vc{k}}|^2$.
Following the same steps as in the previous subsection we find that 
$\gps_k = \gps(t_k)$ and $\tge_k = \tge(t_k)$ are given by
\equ{
e^{-\gm^2 \gps_k} = \frac{1}{||\mx{E}||^2} \frac{N_k}{N_\infty},
\qquad\qquad
\tge_k = \frac{1}{2 N_k}.
}
So as long as $N_\infty \mg N_k$ our assumption of using the asymptotic 
behaviour for $\gps\ra\infty$ at time $t_k$ is very good. Moreover, since in
this limit $\tge$ is the largest of the three slow-roll functions, we see that
the slow-roll approximation is also valid.
Hence we can use the leading order slow-roll estimate of (\ref{DkSR})
for $|D_{\vc{k}}|^2$: 
\equ{
|D_{\vc{k}}|^2 =\frac{\gk^2}{2} \frac{H_k^2}{\tge_k}
= \frac{2}{3} \, \gm^2 \, N_k^2,
}
where, apart from the previous two expressions, we also used (\ref{Nlim}) 
for $H$. This result agrees with (\ref{Dkflatcentpot}) for identical masses.

\subsection{Slow roll on a curved manifold}
\labl{SlowRollCurved}

Now we turn to the slow-roll behaviour of scalar fields that parameterize a curved
manifold that is isotropic around a point. We start with setting up the general
framework, which we clarify by examples and expand upon in the special cases
discussed in the next  subsections.
Consider an $N$-dimensional 
manifold with coordinates $\Bgf$ and metric $\mx{G}(\Bgf)$ given by
\equ{
\mx{G}(\Bgf) = 
g(\Bgf) \left(
\Id_N + \frac {\gl(\Bgf)}{1 - \gl(\Bgf)} \, \mx{Q}
\right),
\labl{metriccurved}
}
with $g(\Bgf)\neq 0$ and $\gl(\Bgf)\neq 1$. The matrix $\mx{Q}$
is the projection operator defined by
\equ{
\mx{Q} \equiv \vc{e}_0^{\;} \vc{e}_0^T,
\qquad\mbox{with}\qquad
\vc{e}_0 \equiv \frac {\Bgf}{\gf}.
\labl{defQ}
}
Here $\gf = \sqrt {\Bgf^T \Bgf}$ represents the coordinate length of 
the vector $\Bgf$, which should not be confused with 
$|\Bgf| = \sqrt{\Bgf^T \mx{G} \Bgf}$.
By taking this form for the metric we have of course restricted ourselves to
manifolds that are isotropic around a point, but
it covers some general, interesting cases, e.g.\ the sphere with either 
embedding (see section \ref{secquadrcurv}) or stereographical coordinates. 
Also, the equations of motion in a central potential obtained with 
this metric are identical to those obtained in the case of a more general 
metric, as we will explain in section \ref{seccentpot}.
The inverse of this metric and the determinant are given by
\equ{
\mx{G}\inv = \frac{1}{g} \left( \Id_N - \gl \mx{Q} \right),
\qquad\qquad
\det \mx{G} = \frac{g^N}{1 - \gl}.
\labl{Ginv}
}
For the determinant we used the relation $\ln\det\mx{G} = \tr\ln\mx{G}$ and the
fact that $\tr\, \mx{Q}=1$.

Inserting our special choice for the metric into the (pure) slow-roll 
equation of motion for $\Bgf$ gives:
\equ{
\Dot \Bgf = - \frac 2{\sqrt{3} \, \gk} \, \mx{G}\inv \Bder^T \sqrt{V(\Bgf)} = 
- \frac 2{\sqrt 3} \frac 1{\gk g} \left(
\Bder^T \sqrt{V}
- \frac {\gl}{\gf^2} ( \Bder \sqrt{V} \; \Bgf ) \Bgf
\right).
\labl{emSRcurved}
}
Notice that $\Bder \sqrt{V} \: \Bgf = \gf^a \, \der_a\sqrt{V}$ is a scalar.
In general this vector equation can be hard to solve, but in practice we often 
have some information from the corresponding flat case that we can use. 
In particular, we can often determine the trajectories that the scalar fields 
follow  through the flat field space. On the other hand, it is much harder to
calculate exactly how the scalar fields move along these trajectories as a
function of time, but this still means that we have reduced the system of $N$
differential equations for $\Bgf$ to a single one that gives the velocity along
the trajectories. In other words, the trajectories of the slow-roll equation 
of motion for the flat case can be written as
\equ{
\Bgf_{\mathrm{flat}}(t) = \vc T(\gps(t), \Bgf_0),
\qquad\mbox{with}\qquad
\vc T(0, \Bgf_0) = \Bgf_0 
\qquad\text{and}\qquad
\gps(t_0) = 0
}
(with $\vc{T}$ a known function), where the function $\gps(t)$ has to satisfy 
the differential equation
\equ{
\dot \gps = - \frac 2{\sqrt 3 \,\gk } 
\frac { \Bder \sqrt V\, \vc T_{,\gps} } { \vc T_{,\gps}^T \vc T^{\;}_{,\gps} }
\qquad\qquad\mbox{(flat case)}.
\labl{paraTraj}
}
An example of this was given in (\ref{ansatz_Bgf}) and (\ref{eqgps}) for the
case of a quadratic potential.

This flat solution can be generalized to curved manifolds with a metric 
of the form introduced above by defining
\equ{
\Bgf_{\mathrm{curved}}(t) = s(\gps(t)) \, \vc T(\gps(t), \Bgf_0).
\labl{ansatz_Bgf_curved}
}
Here $\vc{T}$ is the same function as above, while the
differential equation \eqref{paraTraj} for $\gps$ is slightly modified to 
\equ{
\dot \gps = - \frac 2{\sqrt 3 \gk} \frac 1{g s}  
\frac { \Bder \sqrt V \,\vc T_{,\gps} } { \vc T_{,\gps}^T \vc T^{\;}_{,\gps} }
\qquad\qquad{\mbox{(curved case)}}.
\labl{eqgpscurved}
}
By inserting our ansatz for the solution into the
equation of motion we find that the factor $s(\gps)$ has to satisfy 
\equ{
- \frac {s_{,\gps}}{s} = 
\gl \, \frac {\vc T_{,\gps}^T \vc T^{\;}_{,\gps}}{\vc T^T \vc T} \,
\frac { \Bder \sqrt V \, \vc{T}} {\Bder \sqrt V\, \vc T_{,\gps}}
\qquad\mbox{and}\qquad
s(0) = 1.
\labl{eqscurved}
}
We give examples of this general method in the following subsections.
 
Next we discuss the definition and evaluation of the slow-roll functions 
$\tge$, $\tget^\parallel$ and $\tget^\perp$. 
To this end we define functions $C_n(V)$ as follows:
\equ{
C_1(V) =  \gk^2 \; \frac {2V}{\gf_0^2},
\qquad
C_n(V) = \gk^{2n}\;  
\frac {\Bnabla V (\mx G\inv \Bnabla^T \Bnabla V)^{n-2} 
\mx G\inv \Bnabla^T V}{\gf_0^2},
\quad n \geq 2.
\labl{defC}
}
The functions $C_n(V)$ are more than simply the curved generalization of the 
functions $F_n$: the $C_n$ are defined for an arbitrary potential, while
in the definition of the $F_n$ we have assumed a quadratic potential and made
use of the fact that we can determine the trajectories of the fields in that
case.
Using the Green-Schwarz inequality we can derive the following 
inequalities for positive integers $n, p$ with $0 < p < n$:
\equ{
C_{n}^2 \leq C_{2p} C_{2(n-p)},
}
which follows by writing 
\(
\gf_0^2 C_n = \gk^{2n} {\Bnabla V (\mx G\inv \Bnabla^T \Bnabla V)^{p-1} 
\mx G\inv (\Bnabla^T \Bnabla V \mx G\inv )^{n-p-1}  \Bnabla^T V}.
\)
The slow-roll functions \eqref{slowrollfun2} can be written as
\equ{
\tge = \frac {2}{\gk^2 \gf_0^2} \frac{C_2}{C_1^2},
\qquad
\tget^\parallel = - \frac {2}{\gk^2 \gf_0^2} 
\frac {C_3 C_1 - C_2^2}{C_1^2 C_2}, 
\qquad
(\tget^\perp)^2 = 
\left(
\frac {2}{\gk^2 \gf_0^2} 
\right)^2 
\frac{C_2 C_4 - C_3^2} {C_1^2 C_2^2},
\labl{slowrollcurved}
}
which are the same expressions as in the case of a quadratic potential on  a
flat manifold \eqref{slowrollfun_flat}, but with the $F_n$ replaced by 
$C_n(V)$.  The only inequality for the functions $C_n$ that is directly
applicable is  for $n = 3, p = 1$: $C_3^2 \leq C_2 C_4$, which implies that the
square  $(\tget^\perp)^2$  is positive, as it should be. 

On a curved manifold the second order covariant derivative $\Bnabla^T \Bnabla V$ 
contains connection terms. We now compute what these terms are in the case of
our special metric. It is convenient to work out the special combination
\equ{
\vc v = (\Bnabla^T \Bnabla V) \mx G\inv \Bnabla^T V,
\labl{defv}
}
since that is how the connection enters into the expressions of the slow-roll
functions, as can be seen by writing
\equ{
C_3 = \gk^6 \frac {\Bnabla V \mx G\inv \vc v}{\gf_0^2},
\qquad\qquad
C_4 = \gk^8 \frac {\vc v^T \mx G\inv \vc v}{\gf_0^2}.
}
This vector $\vc v$ can be split into two vectors $\vc{v} = \vc{v}_F +
\vc{v}_G$, given by
\equ{
\vc v_F = (\Bder^T\Bder V) \mx G\inv \Bder^T V,
\qquad
(\vc v_G)_a = - \gG^c_{\; ab} \der_c V (\mx G\inv \Bder^T V)^b 
= \half (\mx G\inv)^{bc}_{\;\;\; ,a}  \der_b V  \der_c V. 
}
Here we have used that 
\(
- \gG^c_{\; ab} w_c (\mx G\inv \vc w)^b = 
- \gG_{c\, ab}  (\mx G\inv \vc w)^b(\mx G\inv \vc w)^c = 
\half (\mx G\inv)^{bc}_{\;\;\; ,a} w_b w_c
\)
for any vector $\vc w$. Next we compute the derivative of the 
inverse metric. Using the definitions of the inverse metric 
$\mx G\inv$ (\ref{Ginv}) and the projection operator $\mx Q$ (\ref{defQ}), 
we obtain 
\equ{
\half (\mx G\inv)^{bc}_{\;\;\; ,a} = 
- \frac {g_{,x}}g \frac {\gf_a}{R^2} ( \mx G\inv)^{bc} 
- \frac {\gl_{,x}}g \frac {\gf_a}{R^2} \mx Q^{bc} 
- \frac {\gl}{2g} \frac 1{\gf^2} \left(
\gd^b_a \gf^c + \gd^c_a \gf^b - 2 \frac {\gf^b\gf^c}{\gf^2} \gf_a
\right).
}
Here we have assumed that the metric functions $g$ and $\gl$ only depend
implicitly on $\Bgf$ via the quantity $x \equiv \gf^2/R^2$, with $R$ the
characteristic radius of curvature of the manifold. 
This leads to the final result for $\vc v_G$:
\equ{
\vc v_G = - \frac{1}{g R^2} \left[
g_{,x} \, \Bder V \mx G\inv \Bder^T V 
+ \frac{\gl}{\gf^2} \Bigl( \frac {\gl_{,x}}{\gl} - \frac{R^2}{\gf^2} \Bigr) 
( \Bder V \, \Bgf)^2 \right] \Bgf
- \frac {\gl}g (\Bder V \Bgf) \frac {\Bder^T V}{\gf^2}.
}
In the next subsections we work out these expressions for the slow-roll
functions in the cases of some special potentials.

\subsubsection{Scalar fields on a curved manifold with a central potential}
\labl{seccentpot}

We consider the case that the potential $V_c(\Bgf)$ is a 
central potential around the origin: it is a function of the coordinate 
length $\gf$ only. The first and second order gradients $\Bder^T V_c$ and 
$\Bder^T \Bder V_c$ are then given by 
\equ{
\Bder^T V_c = V_{c,\gf} \vc{e}_0
\qquad\mbox{and}\qquad
\Bder^T \Bder V_c = V_{c,\gf\gf} \mx Q + \frac 1{\gf}V_{c,\gf} 
( \Id - \mx Q).
}
As in our first example of scalar fields with equal masses on a flat manifold in
section \ref{eqmassflat}, we find that the vector slow-roll equation of motion 
reduces to a scalar equation:
\equ{
\Dot\Bgf = - \frac {1}{\sqrt 3 \gk} \frac{1-\gl}g 
\frac {V_{c,\gf}}{\sqrt V_c} \, \vc{e}_0
\qquad\Rightarrow\qquad
\Dot\gf = - \frac 1{\sqrt 3 \gk} \frac{1-\gl}g 
\frac {V_{c,\gf}}{\sqrt V_c}.
\labl{eqmotcentpotcurv}
}
Notice that if we take a more general metric
\equ{
\tilde{\mx{G}}(\Bgf) = 
g(\Bgf) \left(
\Id_N + \sum_{n = 0}^{N-1}
 \frac {\gl_n(\Bgf)}{1 - \gl_n(\Bgf)} \mx{Q}_n
\right),
}
with a sum of mutually orthogonal projectors $\mx{Q}_n$, instead 
of just the one $\mx{Q} = \mx{Q}_0$, no other terms appear. This is because 
 $\Bder V_c$ is pointing in the radial direction $\vc{e}_0$ and $\vc{e}_0$ 
is an eigenvector of the metric $\tilde{\mx{G}}$ as well as of metric $\mx{G}$: 
\(
\tilde{\mx{G}} \vc{e}_0 = \mx{G} \vc{e}_0 = \frac{g}{1 -\gl} \vc{e}_0.
\)

Before going on to discuss the slow-roll functions to leading order in slow roll,
we note that we can also say something about the exact slow-roll function 
$\tget^\perp$. From the exact definition of $\tilde\Bget$ in (\ref{slowrollfun}) 
and the exact equation of motion (\ref{eqmotbackFriedmann}) we get
\equ{
(\tget^\perp)^2 = 
%\frac{(\cD_t \Dot{\Bgf})^T \mx{G} 
%\mx{P}^\perp \cD_t \Dot{\Bgf}}{H^2 |\Dot{\Bgf}|^2}
%= 
\frac{\Bder V (\Id - \vc{e}_1 \vc{e}_1^T \mx{G}) \mx{G}\inv \Bder^T V}
{H^2 |\Dot{\Bgf}|^2}
= \frac {1- \gl}{g} 
\left( \frac{V_{c,\gf}}{ H |\Dot \Bgf|} \right)^2 
\left( 1 - \cos^2 \ga \right),
\labl{tget_exact}
}
where $\ga$ denotes the angle between the unit vectors 
$\vc{e}_0$ and $\vc{e}_1$, or equivalently, between the position 
vector $\Bgf$ and the velocity $\Dot\Bgf$. 
This angle $\ga$ is defined in curved field space as
\equ{
\cos \ga = \frac{\vc{e}_1 \cdot \vc{e}_0}{| \vc{e}_1|  |\vc{e}_0|} = 
%\frac{\Dot \Bgf \cdot \Bgf}{| \Dot \Bgf |  | \Bgf |} = 
\sqrt{\frac {g} {1 - \gl}} \, \vc{e}_1^T \vc{e}_0.
}
Notice that while $|\vc{e}_1|=1$, this is not the case for $|\vc{e}_0|$, as
$\vc{e}_0$ was defined as having coordinate length equal to one.
From this formula we infer that if the field velocity $\Dot{\Bgf}$ is pointing 
in the same direction (up to orientation) as the coordinate vector $\Bgf$ (i.e.\ 
$\ga = 0, \pi$), the slow-roll function $\tget^\perp$ vanishes. 
Notice that if this happens, it holds for all time, as can be seen from the
exact equation of motion.

Next we work out the expressions for the slow-roll functions to leading order
in slow roll, given in (\ref{slowrollcurved}). As discussed in the previous
section, it is useful to first calculate the vector $\vc{v}=\vc{v}_F +
\vc{v}_G$, defined in  (\ref{defv}) ff.. We find in the case of a central
potential
\equ{
\vc v = 
\frac {1-\gl}g V_{c,\gf} \left(
V_{c,\gf\gf} - \frac {g_{,x}}g \frac {\gf}{R^2} V_{c,\gf} 
\right) \vc{e}_0  
- \frac {\gl_{,x}}g \frac {\gf}{R^2} (V_{c,\gf})^2 \vc{e}_0,
}
which leads to
\equ{
\tge = \frac {1-\gl}{2g} 
\left( \frac {V_{c,\gf}}{\gk V_c} \right)^2, 
\qquad
\tget^\parallel - \tge = 
-  \frac {1-\gl}{ g} 
\left(
\frac {V_{c,\gf\gf}}{\gk^2 V_c} 
-  \frac {g_{,x}}{g} \frac {\gf}{R^2} 
\frac {V_{c,\gf}}{\gk^2 V_c}
\right) 
+ \frac {\gl_{,x}}{g} 
\frac {\gf}{R^2} \frac {V_{c,\gf}}{\gk^2 V_c}.
\labl{slowrollcentpot}
}
Furthermore, $\tget^\perp = 0$, as can most easily be understood by realizing
that the Green-Schwarz inequality is saturated ($C_3^2 = C_2 C_4$) if $\vc{v}$
and $\Bnabla V$ are (anti-)parallel. This is in agreement with the result
(\ref{tget_exact}), since to leading order in slow roll $\cos^2 \ga = 1$.

To illustrate various aspects of the general discussion above, 
we now turn to an example: a quadratic central potential
of scalar fields with identical masses $\gk^{-1} m$, which are the local 
embedding coordinates on an $N$-dimensional sphere with radius $R$. 
These coordinates $\Bgf$ are induced by embedding the sphere in an $N+1$ 
dimensional Euclidean space, so that by construction $\gf^2 = \Bgf^T\Bgf < R^2$. 
At least two of these coordinate systems are needed to cover the whole sphere. 
In the slow roll discussion here, we stay within one coordinate system because the 
quadratic potential is minimal in the origin of this system. 
The metric, its inverse, and the connection are given by
\equ{
\mx G = \Id  + \frac {\gf^2}{R^2 - \gf^2} \, \mx Q, 
\qquad
\mx G^{-1} = \Id  - \frac{\gf^2}{R^2} \, \mx Q,
\qquad\text{and}\qquad
\gG^a_{\; bc} = \frac{\gf^a}{R^2} \, G_{bc}.
\labl{metric_sphere_em}
}
Hence $g=1$ and $\gl = x = \gf^2/R^2 < 1$ in terms of the general metric
(\ref{metriccurved}).

Inserting the relevant quantities into the slow-roll equation of motion
(\ref{eqmotcentpotcurv}) we find
\equ{
\dot{\gf} = - \sqrt{\frac{2}{3}} \, \frac{m}{\gk^2} 
\lh 1 - \frac{\gf^2}{R^2} \rh.
\labl{EmbCoorEM}
}
Solving this equation, multiplying by a constant unit vector, and applying
the initial condition $\Bgf(0) = \Bgf_0$, we get the following answer for the
solution of the background equation to leading order in slow roll:
\equ{
\Bgf(t) 
%= \frac{1 - \frac{R}{\gf_0} \tanh \gg t}
%{1 - \frac{\gf_0}{R} \tanh \gg t} \, \Bgf_0 
= R \, \tanh [\gg (t_\infty - t)] \: \hat{\Bgf_0},
\labl{solcentpot}
}
with $\gg = \sqrt{\frac{2}{3}} \, \frac{m}{\gk^2 R}$ 
and $t_\infty = \frac{1}{2\gg} \ln \frac{R+\gf_0}{R-\gf_0}$.
The slow-roll functions follow immediately from (\ref{slowrollcentpot}):
\equ{
\tge = \frac{2}{\gk^2 R^2}\, \frac {R^2 - \gf_0^2}{\gf_0^2} \, 
\lh \frac{\sinh (\gg t_\infty)}{\sinh [\gg(t_\infty - t)]} \rh^2,
%\frac{R^2 - \gf_0^2}
%{(\gf_0 \cosh \gg t - R \sinh \gg t)^2},
\qquad\qquad
\tget^\parallel = \frac{2}{\gk^2 R^2},
\qquad\qquad
\tget^\perp = 0.
}
Since $R$ is fixed by the model, the slow roll conditions, $\tge$ and 
$\sqrt{\tge}\, \tget^\parallel$ small, are satisfied if $\gf_0$ is such that 
$R^2 - \gf_0^2 \ll R^2$.
From the Hubble parameter $H=\gk\sqrt{V/3}=m\gf/\sqrt{6}$ we derive the 
expression for the number of e-folds: 
\equ{
N(t) = \int_0^t H \d t = N_\infty - \frac {\gk^2 R^2}{2} 
\ln \cosh [\gg(t_\infty - t)],
\labl{Nefoldscentpot}
}
with $N_\infty = - \frac{1}{4} \gk^2 R^2 \ln (1-\gf_0^2/R^2) \geq 0$, since  $0
\leq \gf_0^2 < R^2$. Therefore, the field should start close to the  equator of
the sphere, $R^2 - \gf_0^2 \ll R^2$, to ensure that sufficient inflation is 
obtained. This is compatible with the requirement that the relevant slow roll 
functions ($\tge$ and $\sqrt{\tge}\, \tget^\parallel$) are small initially.  In
the limit $R \rightarrow \infty$ all results agree  with those we found in the
flat case in section \ref{eqmassflat}. Notice that we can also determine the
solution for $\Bgf$ by using our knowledge from the flat case and the method
described in (\ref{ansatz_Bgf_curved}) ff., but in this particular case that is
more complicated.

We finish with the calculation of the expression for the coefficient 
$|D_{\vc{k}}|^2$, following the steps outlined in section \ref{eqmassflat}.
The time $t_k$ and the slow-roll function $\tge$ at $t_k$ are given by
\equ{
\cosh [\gg(t_\infty - t_k)] = \exp\lh\frac{2 N_k}{\gk^2 R^2}\rh
\qquad\Rightarrow\qquad
\tge_k = \frac{2}{\gk^2 R^2} \frac{1}{\exp\lh\frac{4 N_k}{\gk^2 R^2}\rh-1}.
%\sinh(\ga - \gg t_k) = \sqrt{\exp\lh\frac{4 N_k}{\gk^2 R^2}\rh - 1},
%\qquad
%\tanh(\ga - \gg t_k) = \sqrt{1 - \exp\lh - \frac{4 N_k}{\gk^2 R^2}\rh}.
}
This confirms that the slow-roll approximation is still valid at $t_k$. 
For the model dependent factor $|D_{\vc{k}}|^2$ we obtain to leading order 
in slow roll
\equ{
|D_{\vc{k}}|^2 = \frac{2}{3} \, m^2 \, 
\lh \frac{\gk^2 R^2}{2} \sinh \frac{2 N_k}{\gk^2 R^2} \rh^2.
\labl{Dkcentpotcurv}
}
In this calculation we have assumed that $\tget^\parallel$ itself is small, 
which means that $R$ is larger than the Planck radius $\gk\inv$. If this is 
not the case, more terms in \eqref{DkSR} have to be taken into account. 
In the limit that $R \rightarrow \infty$ this result agrees with the flat case
(\ref{Dkflatcentpot}).

\subsubsection{Scalar fields with different masses on a curved manifold}
\labl{secquadrcurv}

Next we consider the case of the quadratic potential $V_2$, defined in 
(\ref{sqpot}), on a curved manifold with metric (\ref{metriccurved}).
The slow-roll equation of motion in this situation reads
\equ{
\Dot \Bgf = - \sqrt{\frac{2}{3}} \frac{1}{g \gk^2 \sqrt {\Bgf ^T \Bsfm^2 \Bgf}}
\left(
\Bsfm^2 - \gl \frac {\Bgf^T \Bsfm^2 \Bgf}{\gf^2} \Id 
\right) \Bgf.
}
Since we know the trajectories of $\Bgf$ in the flat field case
(\ref{ansatz_Bgf}), we can use the method described in
(\ref{ansatz_Bgf_curved}) ff.\ to determine the solution, or at least the
trajectories, of $\Bgf$ in the curved field space. The equation for $s(\gps)$
can be solved analytically for the sphere with  embedding coordinates, which
was also considered in the previous subsection. 

Before turning to this example, we first work out the expressions for the 
functions $C_1(V_2),\ldots,C_4(V_2)$ in terms of the flat functions $F_n$ 
defined in \eqref{defF}. These expressions can be used to determine the
slow-roll functions (\ref{slowrollcurved}). Because we found many properties
and estimates for the $F_n$ in section \ref{flatmasses},  a lot of additional
properties are obtained for the functions $C_n(V_2)$ in this way. To find the
relation between the $C_n(V_2)$ and the $F_n$ it is convenient  to define
intermediate functions $\tF_n$ that incorporate some of the non-flat metric
aspects, but not the full covariant derivatives (connection  terms). They are
defined as follows:
\equ{
\tF_n = 
\frac{\dsp \Bgf^T \mx G (\mx G\inv  \Bsfm^2)^n \Bgf}{\gf_0^2},
\qquad\qquad n \geq 1.
}
Notice that apart from the metric aspects, the $\tF_n$ also contain an extra
factor of $s^2$ as compared to the $F_n$, since 
$\Bgf = s(\gps) e^{-\half \Bsfm^2 \gps} \Bgf_0$ according to
(\ref{ansatz_Bgf_curved}).
The functions $\tF_n$ can be expressed in terms of the $F_n$; for the first 
four functions $\tF_n$ we find by 
inserting the definition (\ref{Ginv}) of $\mx{G}\inv$
\equa{
\tF_1 & = s^2 F_1, 
\qquad\qquad
\tF_2 = \frac{s^2}{g} \left( F_2 - \gL F_1 \right),
\qquad\qquad
\tF_3 =  \frac{s^2}{g^2} \left( F_3 - 2 \gL F_2 + \gL^2 F_ 1 \right),
\non \\
\tF_4 & =  \frac{s^2}{g^3} \left( 
F_4 - \gL\left( 2 F_3 + \frac {F_2^2}{F_1} \right) + 
3 \gL^2 F_2 - \gL^3 F_1
\right),
\qquad\mbox{with}\qquad
\gL = \gl \frac{F_1}{F_0}.
}
Here we used that 
\(
\Bgf^T \Bsfm^{2p} \mx Q \Bsfm^{2q}  \mx Q \ldots  \Bsfm^{2r} \Bgf = 
( \frac{F_p}{F_0}  \frac{F_q}{F_0} \ldots  \frac{F_r}{F_0} ) F_0 s^2. 
\)
Writing the functions $\tF_n$ as 
\(
\tF_n = \Bgf^T ( \Bsfm^2 \mx G\inv)^p \mx G 
(\mx G\inv \Bsfm^2)^{n-p} \Bgf/\gf_0^2,
\)
we obtain the same Green-Schwarz inequalities as for the $C_n$:
\equ{
( \tF_n )^2 \leq \tF_{2p} \tF_{2(n-p)}
}
for integer $0 < p < n$. 
The next (and final) step is to write the functions $C_n(V_2)$ in terms of the
$\tF_n$. It is easy to show that
\equ{
C_1(V_2) = \tF_1 = s^2 F_1
\qquad\text{and}\qquad
C_2(V_2) = \tF_2 = \frac{s^2}{g} \lh F_2 - \gL F_1 \rh.
}
For the functions $C_3$ and $C_4$ we use the vector $\vc v$ defined in 
\eqref{defv}, which in the case of a quadratic potential can be written as
\equ{
\vc v = \frac 1{\gk^4} 
\left(
\Bsfm^2 \mx G\inv \Bsfm^2 - \frac {\gL}g \Bsfm^2 
- \frac {H}g \Id 
\right) \Bgf,
}
where 
\(
H = g_{,x} \frac {\gf_0^2}{R^2} \tF_2 + 
\gL \tF_1 
\left( 
\frac {\gl_{,x}}{\gl} \frac {\gf_0^2}{R^2} - \frac 1{s^2 F_0}
\right).
\)
By inserting this into the expressions for $C_3$ and $C_4$ we obtain
\equ{
C_3(V_2) = \tF_3 - \frac {\gL}g \tF_2 - \frac {1-\gl}g \frac Hg \tF_1,
\\
C_4(V_2) = \tF_4 - 2 \frac {\gL}g \tF_3  + \Bigl(\frac {\gL}g \Bigr)^2 \tF_2 
- 2 \frac {1 -\gl}g  \frac Hg 
\left[
\tF_2 - \frac {\gL}g \tF_1 - \frac H{2g} \, s^2 F_0 
\right].
\non
}

Next we discuss the example mentioned above. 
We consider the case where the manifold on which the fields live is a sphere
with radius $R$. We again use embedding coordinates, i.e.\ $g=1$ and 
$\gl=x = \gf^2/R^2$. Solving equation (\ref{eqscurved}) for $s$ we find 
\equ{
\frac{s_{,\gps}}{s^3} = \frac{\gf_0^2}{2 R^2} \, F_1(\gps) 
\qquad\Rightarrow\qquad
\frac{1}{s^2(\gps)} = 1 - \frac{\gf_0^2}{R^2} \lh 1 - F_0(\gps) \rh.
}
The trajectories of $\Bgf$ and the differential equation for $\gps$
(\ref{eqgpscurved}) are given by
\equ{
\Bgf(\gps) = s(\gps) e^{-\half \Bsfm^2 \gps} \Bgf_0,
\qquad\qquad
\dot{\gps} = \sqrt{\frac{2}{3}} \frac{2}{\gk^2 \gf_0} \frac{1}{s(\gps)}
\frac{1}{\sqrt{F_1(\gps)}}.
}
As in the flat case, $\gps$ is monotonously increasing, starting at zero on
$t=0$, and reaching $\infty$ when $\Bgf=0$.
The slow-roll functions are again determined from (\ref{slowrollcurved}), using
the expressions for the $C_n$ derived above. However, since the resulting 
expressions are quite large, we only give $\tge$ here:
\equ{
\tge \ = \ \frac{2}{\gk^2 \gf_0^2} \frac{1}{s^2} \frac{F_2}{F_1^2} 
- \frac{2}{\gk^2 R^2}
\ = \ \frac{2}{\gk^2 \gf_0^2} \lh 1 - \frac{\gf_0^2}{R^2} \rh
\frac{F_2}{F_1^2}
+ \frac{2}{\gk^2 R^2} \frac{F_0 F_2 - F_1^2}{F_1^2},
}
which, using the Green-Schwarz inequality (\ref{GrSchwF}), is seen to be larger 
than or equal to zero, as it should. For the number of e-folds we find
\equ{
N % = \int_0^t H \d t
= \frac{\gk^2 \gf_0^2}{4} \int_0^\gps s^2 F_1 \d\gps
= N_\infty - \frac{1}{4} \gk^2 R^2 \ln \lh 1 + \frac{\gf_0^2}{R^2 - \gf_0^2} \, 
F_0(\gps) \rh,
}
where $N_\infty$ is the same as in the previous subsection:
$N_\infty = \frac{1}{4} \gk^2 R^2 \ln(R^2/(R^2-\gf_0^2))$.

As mentioned above the functions $F_n$ satisfy the estimates
(\ref{estF}). This means that we can give the following estimates for $H$,
$\tge$ and $N$ in the limit $\gps\ra\infty$, using that $\gf_0 < R$:
\equa{
H & \ra \frac{\gf_0 \mu ||\mx{E}||}{\sqrt{6}} 
\left[ 1 - \frac{\gf_0^2}{R^2} \lh 1 - ||\mx{E}||^2 e^{-\mu^2 \gps} \rh
\right]^{-\half} e^{-\half \mu^2 \gps},
\qquad
\tge \ra \frac{2}{\gk^2 \gf_0^2} \frac{1}{||\mx{E}||^2} 
\frac{R^2 - \gf_0^2}{R^2} \, e^{\mu^2 \gps},
%+ \frac{2 \chi}{\gk^2 R^2} (\gr-1)^2 e^{-(\gr-1)\mu^2 \gps},
\non\\
N & \ra N_\infty - \frac{1}{4} \gk^2 R^2 \ln \lh 1 + \frac{\gf_0^2}{R^2 - \gf_0^2}
||\mx{E}||^2 e^{-\mu^2 \gps} \rh.
}
Following the same steps as in the previous subsection to determine
$|D_{\vc{k}}|^2$, we find for the critical value $\gps_k$
\equ{
e^{-\mu^2 \gps_k} = \frac{1}{||\mx{E}||^2} \frac{R^2 - \gf_0^2}{\gf_0^2}
\lh \exp \lh \frac{4 N_k}{\gk^2 R^2} \rh - 1 \rh
= \frac{1}{||\mx{E}||^2} \frac{\exp \lh \frac{4 N_k}{\gk^2 R^2} \rh - 1}
{\exp \lh \frac{4 N_\infty}{\gk^2 R^2} \rh - 1}.
}
Hence, as in the flat case, the assumption of looking at the asymptotic
behaviour for $\gps$ at time $t_k$ is a good approximation 
if $N_\infty \mg N_k$.
Finally, the expression for $|D_{\vc{k}}|^2$ in this limit is
\equ{
|D_{\vc{k}}|^2 = \frac{2}{3} \, \mu^2\, 
\lh \frac{\gk^2 R^2}{2}
\, \sinh \frac{2 N_k}{\gk^2 R^2} \rh^2,
}
in agreement with (\ref{Dkcentpotcurv}) for identical masses.

\np

\section{Conclusions}

In this paper we have analyzed scalar gravitational perturbations on 
a Robertson-Walker background in the presence of multiple scalar fields 
that take values on a (geometrically non-trivial) field manifold during 
slow-roll inflation. 

We have modified the definitions of the well-known slow-roll parameters  to
define slow-roll functions in terms of the Hubble parameter, background field
velocity and their  derivatives in the case of multiple scalar field
inflation. This means  that the slow-roll functions have become vectors, except
for $\tge$ which is a derivative of the Hubble parameter. Like other relevant
vectors they are split into a component parallel to the scalar field velocity,
and components perpendicular to this velocity vector. To define  the
slow-roll functions we do not need to make the assumption that slow roll 
is valid, but if it is valid one can expand in these functions, giving  the 
relative importance of terms in various equations.
%In contrast, the conventional definitions of the slow-roll 
%parameters in terms of the scalar field potential are obtained as consistency 
%checks by inserting the slow-roll approximated equation of motion back into 
%the definition of the slow-roll functions. In addition, our definitions make 
%it possible to expand in the slow-roll functions, giving  the relative
%importance of terms in various equations. 
For example, if   ${\tge}$,
$\sqrt{\tge}\, \tget^\perp$  and $\sqrt{\tge}\, \tget^\parallel$  are small,
the background equation of motion for the scalar fields can be  approximated
by the (pure) slow-roll equation. 

We set up the combined system of gravitational and matter perturbations in a
way analogous to Mukhanov et al.\ \cite{Mukhanovetal}, but including multiple
scalar fields and effects of a non-trivial field geometry.  
The component of the scalar field perturbations parallel to the background 
field velocity can be eliminated. The remaining perpendicular 
components of the field perturbations and the gravitational potential are 
described by coupled differential equations (\ref{matrixeq}). 
However, the gravitational potential decouples from the perpendicular field 
perturbations if effects of the order of $\sqrt{\tge}\, \tget^\perp$ can be 
neglected; to the same order the background equations reduce to the 
slow-roll equations. 

Since to first order in slow roll the equation for the gravitational potential
is equal to the single field case, it has the same expressions for the solution
as \cite{Mukhanovetal} in the small and long wavelength limits.  Using the
slow-roll functions the corrections due to the transition region between these 
two limits can be estimated. It follows that simply joining the two solutions 
together yields a good approximation during the complete inflationary period,
in the sense that corrections are of higher order in slow roll.    The
transition between the two solutions happens when   $k^2 = |\gth''/\gth|
\approx \cH^2(2\tge+\tget^\parallel)$. Even though  this is unequal to the time
of horizon crossing $k^2 = \cH^2$ of a scale $k$, the corrections to the
gravitational potential if one would use this time instead are suppressed in
slow roll.

The quantum two-point correlation function of the gravitational potential  is
related to the temperature fluctuations that are observed in the CMBR.  The
only physical degrees of freedom that can be quantized in the  system we
consider are the scalar field perturbations. (Only the graviton  states are
physical degrees of freedom on the gravitational side. However,   they are
decoupled because the Einstein equations have been linearized.)  Although the
quantization at the beginning of inflation  involves the scalar field
perturbations, after the Fock space has been  constructed  the time evolution
of the correlator can be calculated as if it is a classical quantity. Choosing
the vacuum as the initial state of inflation is a  good assumption for the
scales that are observable in the CMBR,  since their momenta in the initial
stages of inflation are  much larger than the Planck energy. Taking a thermal
state with the Planck temperature at the beginning of inflation as a first
attempt at improving on the assumption of a pure vacuum leads to corrections to
the vacuum correlator of the order of only $10^{-3}$ or less. Here we assumed
that at least a few e-folds of inflation  occured between  the Planck time and
the moment when the observable scales went through the  horizon. The
gravitational correlation function contains one directly model  dependent
factor that can be expressed in terms of the slow-roll functions.

Finally, we discussed some multiple field examples to illustrate some 
dynamical aspects of the background and compute this model dependent 
factor. On flat manifolds with central or quadratic potentials
the trajectories of the fields in field space can be found in terms of one
function $\gps(t)$, so that only a single differential equation remains. This
equation of motion, as well as many other relevant quantities like the
slow-roll functions and the Hubble parameter, can be expressed in 
functions $F_n$ of this $\gps$. These functions have many useful 
properties which make it possible to analyze slow-roll phenomena 
without having to explicitly solve the (complicated) equation for $\gps$.
They satisfy inequalities that follow from the Green-Schwarz inequality.
Their asymptotic behaviour for large $\gps$ can be used to analyze 
various important quantities when observable scales go through the 
horizon, provided that there have been many e-folds of inflation before 
that. Using that the derivative of $F_n$ equals $-F_{n+1}$ makes it 
possible to integrate the Hubble parameter and obtain the number of e-folds 
as a function of $\gps$.

The generalizations $C_n$ of the functions $F_n$ to curved field space  and to
an arbitrary potential lead to the same expressions for the slow-roll 
functions as in the flat case with a quadratic potential. For such a potential 
on a manifold isotropic around the origin we expressed the $C_n$  in terms of
their flat relatives. If the trajectories of the fields in the flat field
space  are known for a given potential, the corresponding trajectories on  an
isotropic manifold are the trajectories in flat space multiplied by a scalar 
function $s$ of the parameter $\gps$. The sphere with embedding coordinates  is
a special example of an isotropic manifold. The background equations  can be
solved explicitly as a function of time for a quadratic potential with  all
masses equal. If not all masses are equal, it is still possible to find an 
integrated expression for the number of e-folds in terms of $F_0(\gps)$.  For
the flat space examples we considered, the only possibility to obtain a  large
total number of e-folds is by taking large initial field values.  The radius of
curvature of a curved manifold is an additional parameter  that influences the
total number of e-folds: for the examples of  the sphere this number becomes
large  if the initial field values are of the same order as the radius of the
sphere. 
\\[2ex]
{\bf Acknowledgements} 
\\[2ex]
We thank D.J.\ Schwarz for pointing out some subtleties related to the 
transition region where the gravitational potential changes its behaviour. 
This work is supported by the European Commission RTN
programme HPRN-CT-2000-00131.
S.G.N.\ is also supported by  
piority grant 1096 of Deutsche Forschungsgemeinschaft
and European Commission RTN programme HPRN-CT-2000-00152.

\np

\appendix

\section{Derivation of the equation of motion for 
the perpendicular field perturbations}
\labl{derivation}

This appendix is devoted to the derivation of the equation of motion 
for the perpendicular field components 
$\Bgd\vc v = \smash{\frac a{\gk^2 |\Bgf'|}} \Bgd \Bgf^\perp$. 
We start by taking the perpendicular projection of equation 
(\ref{eqmotpert2}). Using the antisymmetric properties of the Riemann 
tensor and the fact that we contract with two identical vectors $\Bgf'$ 
we conclude that
\(
\mx{P}^\perp \mx{R}(\Bgf',\Bgf')\Bgd\Bgf =
\mx{R}(\Bgf',\Bgf')\Bgd\Bgf^\perp. 
\)
The perpendicularly projected equation of motion for $\Bgd\Bgf$ can 
be written as
\begin{multline}
\mx{P}^\perp \lh \cD_\eta^2 + 2 \cH \cD_\eta 
+ a^2 \mx{M}^2 \rh \Bgd\Bgf^\parallel
+ 2 a^2 \gF \mx P^\perp \mx G\inv \Bnabla^T V + 
\mx P^\perp \lh \cD_\eta^2 + 2 \cH \cD_\eta  \rh \Bgd\Bgf^\perp \\
+ \lh a^2 (\mx M^2)^{\perp\perp} - \gD -  \mx{R}(\Bgf',\Bgf') \rh 
\Bgd\Bgf^\perp = 0.
\labl{eqmotpert2proj}
\end{multline}
Here we split $\Bgd\Bgf = \Bgd\Bgf^\parallel + \Bgd\Bgf^\perp$ 
into components parallel and perpendicular to $\Bgf'$.  Furthermore,  
we used that the projection operator commutes with $\gD$ and 
employed the notation defined in \eqref{vecmatproj}.
%\(
%(\mx{M}^2)^{\perp\perp} = \mx{P}^\perp \mx{M}^2 \mx{P}^\perp.
%\)
The last term can be trivially rewritten, so we now derive expressions for 
the first three terms on the left-hand side of \eqref{eqmotpert2proj} in
terms of $u$ and $\Bgd \vc v$ only.

We consider the first term in \eqref{eqmotpert2proj}. 
This term does not vanish in general, since the derivatives and mass matrix 
generally change the direction of $\Bgd\Bgf^\parallel$.
The covariant derivative $\cD_\get^2$ applied to the non-contracted vector 
$\Bgf'$ in
\(
\smash{\Bgd\Bgf^\parallel = 
\frac {\Bgf'\cdot \Bgd\Bgf}{|\Bgf'|^2} \Bgf'}
\)
is canceled by the mass term. This can be seen from the covariantly 
differentiated background equation of motion \eqref{eqmotback2}, 
\equ{
\cD^2_\eta \Bgf' + 2 \lh \cH' - 2 \cH^2 \rh \Bgf' + a^2 \mx{M}^2 \Bgf'
= 0,
\labl{eqmotback2der}
}
where we used $\cD_\eta (\mx G\inv \Bnabla^T V) = \mx{M}^2 \Bgf'$, if we
apply the perpendicular projection operator:
$\mx P^\perp \cD_\get^2 \Bgf' + \mx P^\perp (a^2 \mx M^2 \Bgf') = 0$. 
The remaining contribution of the first term of \eqref{eqmotpert2proj} 
can be written as
\equ{
\mx{P}^\perp \lh \cD_\eta^2 + 2 \cH \cD_\eta + a^2 \mx{M}^2 \rh
\Bgd\Bgf^\parallel
= 
\frac{\gk^2 |\Bgf'|}{a} \;\frac {2}{\gk^2} 
\left [ a \, \frac{\Bgf' \cdot \Bgd\Bgf}{|\Bgf'|^2} \right ]' \;
\frac {(\cD_\eta \Bgf')^\perp}{|\Bgf'|}.
\labl{Pperpphiparal}
}
The factor involving $\Bgf' \cdot \Bgd\Bgf$ can be rewritten in terms of $u$
using  the (0i)-component of the Einstein equation (\ref{Einstein0i2}):
\equ{
\frac{2}{\gk^2} 
\left [ a \, \frac{\Bgf' \cdot \Bgd\Bgf}{|\Bgf'|^2} \right]' 
= 
\frac {2}{\gk^2} \left[ \frac {2\gk}{| \Bgf'|} 
\Bigl( u' + \frac {|\Bgf'|'}{|\Bgf'|} u \Bigr) \right]' 
= 
\frac {4}{\cH\sqrt{2\tge}} 
\left[ u'' + 
\left( \frac{|\Bgf'|''}{|\Bgf'|} 
- 2 \Bigl( \frac{|\Bgf'|'}{|\Bgf'|} \Bigr)^2 \right) u
\right],  
}
where we have used that $\gk |\Bgf'| = \cH \sqrt{2 \tge}$, 
which follows from \eqref{Einstback2} and \eqref{slowrollrel}. 

For the second term in \eqref{eqmotpert2proj} we use the 
background field equation \eqref{eqmotback2} and obtain 
\equ{
2a^2\Phi\, \mx P^\perp \mx G\inv \Bnabla^T V = 
-2\Phi \, (\cD_\eta\Bgf')^\perp = 
\frac {\gk^2 |\Bgf'|}a \;
\frac{4}{\cH \sqrt{2\tge}} (\cH' - \cH^2) u \;
\frac {(\cD_\eta \Bgf')^\perp}{|\Bgf'|},
\labl{term2}
}
using another, but equivalent expression for $\gk|\Bgf'|$: 
$\gk |\Bgf'| = 2 (\cH^2 - \cH')/(\cH\sqrt{2\tge})$. 
Combining these expressions and using the equation of motion for $u$
\eqref{equstep} we find that the first two terms of \eqref{eqmotpert2proj} can
be written as
\equ{
\mx{P}^\perp \lh \ldots \rh \Bgd\Bgf^\parallel
+ 2 a^2 \gF \mx P^\perp \mx G\inv \Bnabla^T V
= 4 \, \frac {\gk^2 |\Bgf'|}a \left [
\frac{\gD u}{\cH \sqrt{2 \tge}}
+ \frac{(\cD_\eta \Bgf')^\perp}{|\Bgf'|} \cdot \Bgd\vc{v} 
\right ] 
\frac{(\cD_\eta \Bgf')^\perp}{|\Bgf'|}.
\labl{term1en2}
}

Moving to the third term of  \eqref{eqmotpert2proj}, we first 
note that substitution of $\Bgd\Bgf^\perp = 
\smash{\frac {\gk^2 |\Bgf'|}a \Bgd\vc v}$ gives
\equ{
\mx P^\perp ( \cD_\get^2 + 2 \cH \cD_\get) \Bgd \Bgf^\perp = 
\frac{\gk^2 |\Bgf'|}a \;
( \Id - \mx P^\parallel ) 
\left[
\cD_\get^2   +
2 \frac{|\Bgf'|'}{|\Bgf'|} \cD_\get  + 
 \frac{|\Bgf'|''}{|\Bgf'|} - \cH^2 - \cH'
\right] \Bgd \vc v.
\labl{term3}
}
Since $\mx P^\parallel$ only gives a non-vanishing contribution 
if it is applied to $\cD_\get \Bgd\vc v$ or $\cD^2_\get \Bgd \vc v$, 
we consider those terms separately and find 
\equ{
\mx P^\parallel 
\left(\cD_\get^2   + 2 \frac{|\Bgf'|'}{|\Bgf'|} \cD_\get \right) \Bgd \vc v = 
- \left(
\frac{(\cD_\eta^2 \Bgf')^\perp}{|\Bgf'|} \cdot \Bgd\vc v + 
2 \frac{(\cD_\eta \Bgf')^\perp}{|\Bgf'|} \cdot \cD_\get \Bgd\vc v
\right) \vc e_1. 
\labl{term3a}
}
To obtain this expression, we wrote 
$\mx P^\parallel = \vc e_1 \Bgf^{\prime \dag}/|\Bgf'|$ and used the 
two relations that are obtained by taking the first and second derivative 
with respect to conformal time of the equation $\Bgf' \cdot \Bgd \vc v = 0$:
\equ{
\Bgf' \cdot \cD_\eta \Bgd\vc v = - \cD_\eta \Bgf' \cdot 
\Bgd\vc v,
\qquad
\Bgf' \cdot \cD_\eta^2 \Bgd\vc v = 
- \cD_\eta^2 \Bgf' \cdot \Bgd\vc v - 
2 (\cD_\eta \Bgf')^\perp \cdot 
\lh \cD_\eta - \frac{|\Bgf'|'}{|\Bgf'|} \rh \Bgd\vc v,
\non
}
where in the second relation we used
\(
\cD_\eta \Bgf' = (\cD_\eta \Bgf')^\perp + 
{\frac{|\Bgf'|'}{|\Bgf'|}} \Bgf'
\)
and the first relation.

Now we can insert the expressions \eqref{term1en2}, 
\eqref{term3} and \eqref{term3a} into \eqref{eqmotpert2proj} 
and find
\equa{
& \left [ \cD_\eta^2 + 
2  \frac{|\Bgf'|'}{|\Bgf'|} \cD_\eta + 
\lh  \frac{|\Bgf'|''}{|\Bgf'|}- 
{\cH'} - \cH^2 \rh 
+ a^2 (\mx{M}^2)^{\perp\perp} - \gD - \mx{R}(\Bgf',\Bgf') 
\right ] \Bgd\vc{v}
\labl{eqdeltav} \\
& + 4 \left [
\frac{1}{\cH\sqrt{2 \tge}} \, \gD u
+ \frac{(\cD_\eta \Bgf')^\perp}{ |\Bgf'|} \cdot \Bgd\vc{v} 
\right ] 
\frac{(\cD_\eta \Bgf')^\perp}{|\Bgf'|}
+ \left [ \frac{(\cD_\eta^2 \Bgf')^\perp}{ |\Bgf'|} \cdot 
\Bgd\vc{v} + 2 \frac{(\cD_\eta \Bgf')^\perp}{ |\Bgf'|} \cdot 
\cD_\eta \Bgd\vc{v} \right ] \vc{e}_1 = 0.
\non
}
Using \eqref{slowrollrel} we obtain the final form in terms of the slow-roll
functions:
\begin{multline}
\left [ 
\cD_\eta^2 + 
2 \cH \lh 1+\tget^\parallel \rh \cD_\eta 
+ \cH^2 \lh 3 \tget^\parallel + (\tget^\perp)^2 + \tgx^\parallel \rh
+ a^2 (\mx{M}^2)^{\perp\perp} - \gD - \mx{R}(\Bgf',\Bgf') 
\right ] \Bgd\vc{v}
\\
+ \left [ 
\cH^2 ( 3 \tilde\Bget + \tilde \Bgx)^\perp \cdot  \Bgd\vc{v} + 
2 \cH \tilde \Bget^\perp \cdot 
\cD_\eta \Bgd\vc{v} \right ] \vc{e}_1 
+ 4 \Bigl[
\cH^2 \tilde \Bget^\perp \cdot \Bgd\vc{v} 
+ \frac{ 1}{\sqrt{2 \tge}} \, \gD u
\Bigr] 
\tilde \Bget^\perp 
= 0,
\labl{eqapp}
\end{multline}
which corresponds to the $\Bgd\vc v$ component of the matrix 
equation \eqref{matrixeq}.

\np

\end{document}